\documentclass[a4paper,fleqn,usenatbib]{mnras}

\usepackage{newtxtext,newtxmath}
\usepackage[T1]{fontenc}
\usepackage{ae,aecompl}

\usepackage{graphicx}	% Including figure files
\usepackage{amsmath}	% Advanced maths commands
\usepackage{amssymb}	% Extra maths symbols
\usepackage{booktabs}
\usepackage{subfig}
\usepackage{float}
\usepackage{xcolor}
\usepackage{multirow}

\newcommand{\GG}[1]{}

\title[Chromodynamical Analysis of S0 Galaxies]{Chromodynamical Analysis of Lenticular Galaxies using Globular Clusters and Planetary Nebulae}

\author[Zanatta et al.]{
Emilio J. B. Zanatta,$^{1}$\thanks{E-mail: emiliojbzanatta@ufrgs.br}
Arianna Cortesi,$^{2}$
Ana L. Chies-Santos,$^{1}$ Duncan A. Forbes,$^{3}$ \newauthor Aaron J. Romanowsky $^{4,5}$, Adebusola B. Alabi,$^{5}$, Lodovico Coccato $^{6}$, \newauthor Claudia Mendes de Oliveira,$^{2}$, Jean P. Brodie,$^{5}$ and Michael Merrifield$^{7}$.
\\
$^{1}$Departamento de Astronomia, Instituto de F{\'i}sica, Universidade Federal do Rio Grande do Sul, Porto Alegre, R.S., Brazil\\
$^{2}$Departamento de Astronomia, Instituto de Astronomia, Geof{\'i}sica e Ciencias Atmosfericas da USP,\\
Cidade Universitaria, CEP:05508900, Sao Paulo, SP, Brazil\\
$^{3}$Centre for Astrophysics \& Supercomputing, Swinburne University, Hawthorn VIC 3122, Australia\\
$^{4}$Department of Physics and Astronomy, San Jos\'e State University, San Jose, CA 95192, USA\\
$^{5}$University of California Observatories, 1156 High St., Santa Cruz, CA 95064, USA\\
$^{6}$European Southern Observatory, Karl-Schwarzschild-Strasse 2, D-85748 Garching bei Munchen, Germany\\
$^{7}$School of Physics and Astronomy, University of Nottingham, University Park, Nottingham NG7 2RD, UK 
}

\date{Accepted 2018 June 28. Received 2018 June 23; in original form 2018 May 7}

\pubyear{2018}

\begin{document}
\label{firstpage}
\pagerange{\pageref{firstpage}--\pageref{lastpage}}
\maketitle

\begin{abstract}

Recovering the origins of lenticular galaxies can shed light on the understanding of galaxy formation and evolution, since they present properties that can be found in both elliptical and spiral galaxies. In this work we study the kinematics of the globular cluster (GC) systems of three lenticular galaxies located in low density environments (NGC\,2768, NGC\,3115 and NGC\,7457), and compare them with the kinematics of planetary nebulae (PNe). The PNe and GC data come from the Planetary Nebulae Spectrograph and the SLUGGS Surveys. 
Through photometric spheroid-disc decomposition and PNe kinematics we find the probability for a given GC to belong to either the spheroid or the disc of its host galaxy or be rejected from the model.
We find that there is no correlation between the components that the GCs are likely to belong to and their colours. Particularly, for NGC\,2768 we find that its red GCs display rotation preferentially at inner radii ($Re<1$). In the case of the GC system of NGC\,3115 we find a group of GCs with similar kinematics that are not likely to belong to either its spheroid nor disc. For NGC\,7457 we find that $70\%$ of its GCs are likely to belong to the disc. Overall, our results suggest that these galaxies assembled into S0s through different evolutionary paths. Mergers seem to have been very important for NGC\,2768 and NGC\,3115 while NGC\,7457 is more likely to have experienced secular evolution.

\end{abstract}


%%%%%%%%%%%%%%%%%%%%%%%%%%%%%%%%%%%%%%%%%%%%%%%%%%

%%%%%%%%%%%%%%%%% BODY OF PAPER %%%%%%%%%%%%%%%%%%

\section{Introduction}

The morphological evolution of galaxies is an interesting and challenging topic in contemporary astrophysics, where lenticular (S0) galaxies play a central role. The morphology-density relation \citep{dress80, dress97, goto03, postman05} and the increase of the number of S0 galaxies at lower redshift, at the expense of the decrease in the number of spiral galaxies in clusters of galaxies \citep{dress83, whitmore93}, led to the idea that S0 galaxies might be the evolutionary products of the transformation of spiral galaxies \citep{larson80}. Such evolution would have been driven by environmental effects such as ram pressure stripping, galaxy harassment, mergers, starvation and alike \citep{gunn72, abadi99, boselli06}. Lenticular galaxies, however, are present in all environments (cluster, groups and field) \citep{van09}. This widens the evolutionary possibilities for creating a lenticular galaxy and raises questions about their uniqueness as a class: could they be the single final product of a variety of events \citep{gunn72, byrd90, quilis00, bekki, bournaurd, ara2008}?

One proposed scenario is that S0 galaxies may form from spiral galaxies which underwent some form of environment-related process and lost most of their gas and spiral structures. These processes therefore would leave behind a disc structure alongside the spheroidal structure of the bulge and halo \citep{gunn72, ara2008, quilis00, byrd90, bekki, bournaurd}. Such scenario would leave the stellar kinematics of the resulting S0 galaxy almost unperturbed, in comparison with the kinematics of the progenitor spiral galaxy \citep{aragon06, buta10, c13b}. On the other hand, S0 galaxies have been suggested to also be formed from mergers between galaxies of unequal mass, which would cause an increase in random motions in the stars \citep{borlaff14, bournaurd, querjeta17}. \citet{bekki} have shown through simulations that merger events that are able to produce flattened early-type galaxies also are likely to affect the structure and kinematics of the globular cluster (GC) systems associated with such galaxies. Moreover, clumpy disc formation is proposed as a way to form galaxy discs through successive mergers of star-forming clumps \citep{vandenbergh96, elme05, shapiro10, genzel11, ceverino15, garland15, fisher17}. In fact, simulations such as the ones from \citet{shapiro10} and \citet{fisher17} show that the formation of GCs in such condition would happen around $z\approx$1-3. Therefore such mechanism could be detected through the kinematics of GCs which would acquire rotation and display disc-like kinematics, even in the halo \citep{inoue13}.

About 90\% of the total mass and angular momentum of a galaxy resides outside one effective radius $(R_{e})$ \citep{roma12}.
By exploring the outer regions of galaxies one may find imprints of their merger histories, and is therefore able to investigate the processes responsible for the build-up of its present mass. 
However, due to the low surface brightness, galaxy outskirts are expensive to probe. Fortunately, tracers such as globular clusters (GCs) and planetary nebulae (PNe) stand as ideal proxies for the study of the kinematics of such regions \citep{hui95, coccato09, forbes12, brodie14}. 

GCs and PNe can be detected out to 5-10 $R_{e}$ in early-type galaxies.
PNe present a strong emission in the [OIII] line (5007 \AA), which allows one to detect them in regions where the galaxy starlight is very faint. The Planetary Nebulae Spectrograph (PN.S, \citet{pns}) observed several early-type galaxies, creating catalogues with an average number of over a hundred PNe per galaxy. As shown in \citet{c11}, this number is enough to recover the stellar kinematics of the galaxy disc and spheroid, since PNe are reliable tracers of the global stellar population of their host galaxy \citep{coccato09, c11}. 
Globular clusters are ubiquitous in early-type galaxies and in the past two decades GC radial velocity catalogues have been acquired for several systems with multi-object spectroscopy in 10m class telescopes (eg. \citet{hanes01}, \citet{cote03}, \citet{hwang08}, \citet{Park12}, \citet{pota13} and \citet{forbes17}).

The present work  analyses the GC systems of three lenticular galaxies in low density environments (NGC\,2768, NGC\,3115 and NGC\,7457) following the method presented in \citet{c16} for the GC system of NGC\,1023. Such study used PNe kinematics derived in \citet{c11, c13b}.

This paper is structured as follows: in section \ref{sec:data} we present an overview of the observational data used in this work and the individual galaxies' general properties.

In section \ref{sec:methods} we briefly discuss the method used for probing GC kinematics using \textsc{GALFIT} \citep{peng02}, maximum likelihood estimation (MLE) and the kinematics of PNe. %originally published in \citet*{c13b}.
In section \ref{sec:results} we present the results on GC kinematics, in addition to considerations on 1D phase-space diagrams and the radial distribution of GCs and PNe for our sample galaxies. In section \ref{sec:discussion} we discuss the results in comparison with the evolutionary paths proposed in the literature for lenticular galaxies. Summary and conclusions are shown in section \ref{sec:conclusion}.

\section{Data}

\label{sec:data}

\begin{figure*}
    \centering
    \subfloat[NGC\,2768]{{\includegraphics[width=0.31\textwidth]{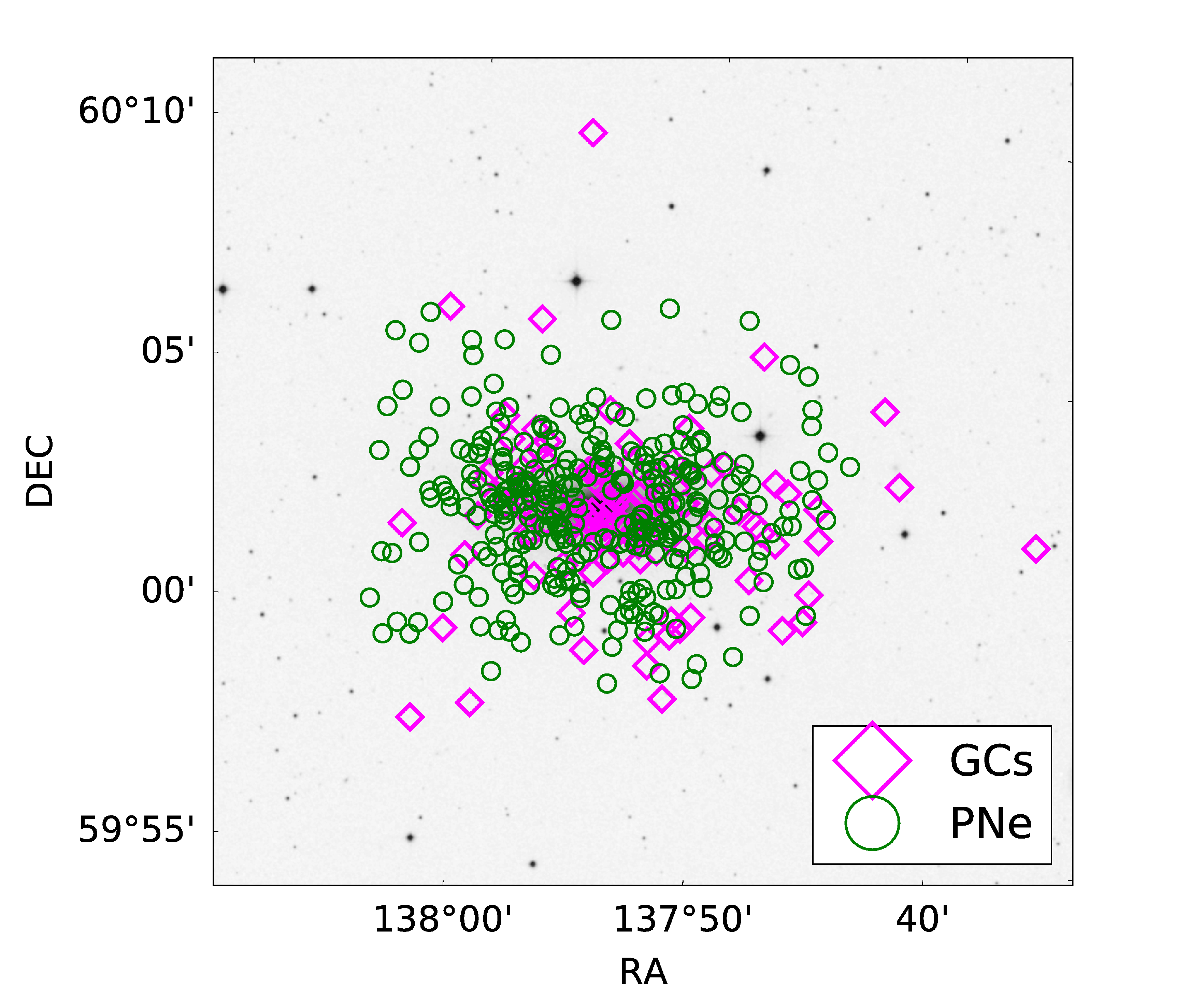} }}
    \subfloat[NGC\,3115]{{\includegraphics[width=0.28\textwidth]{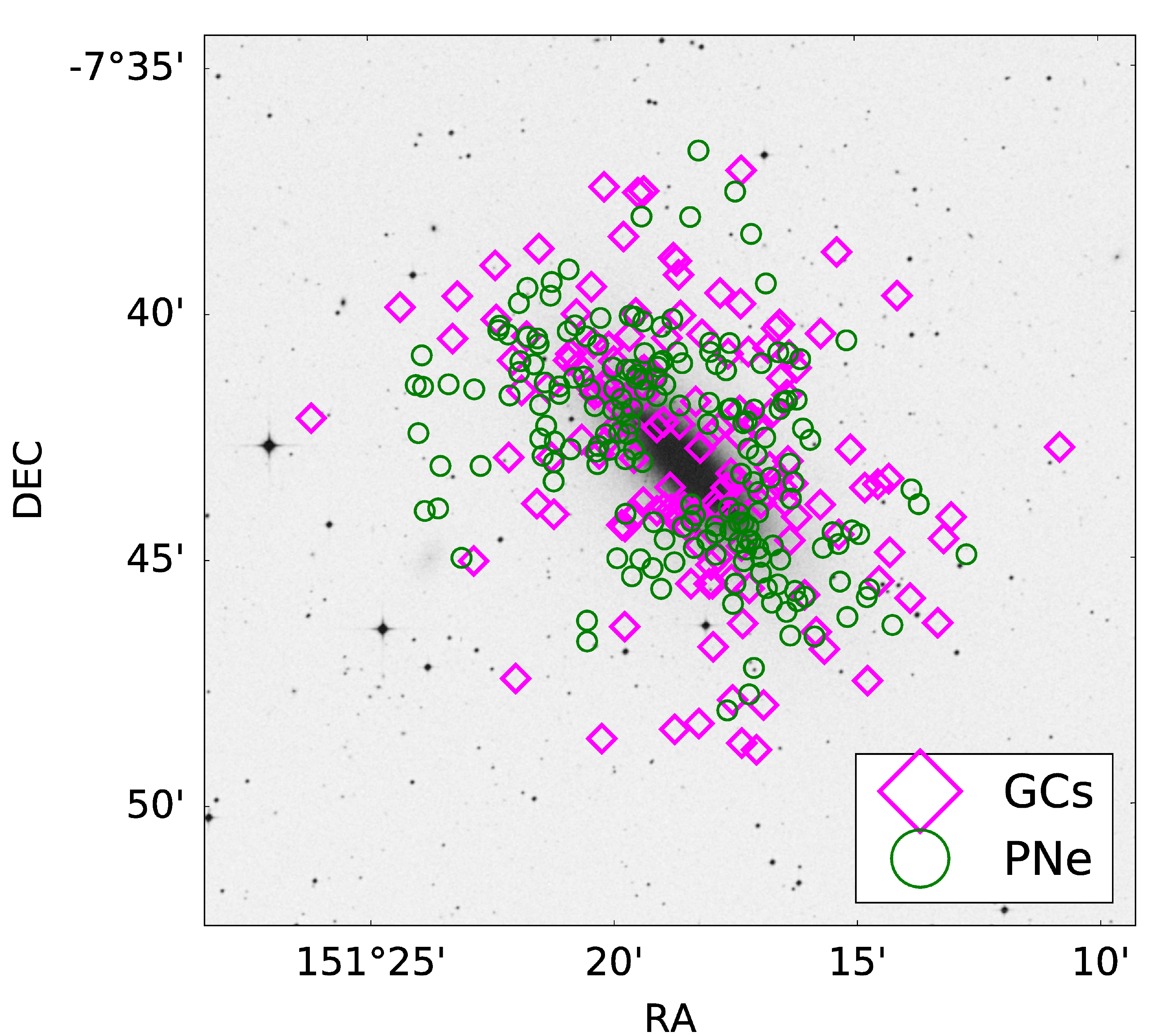} }}
    \subfloat[NGC\,7457]{{\includegraphics[width=0.29\textwidth]{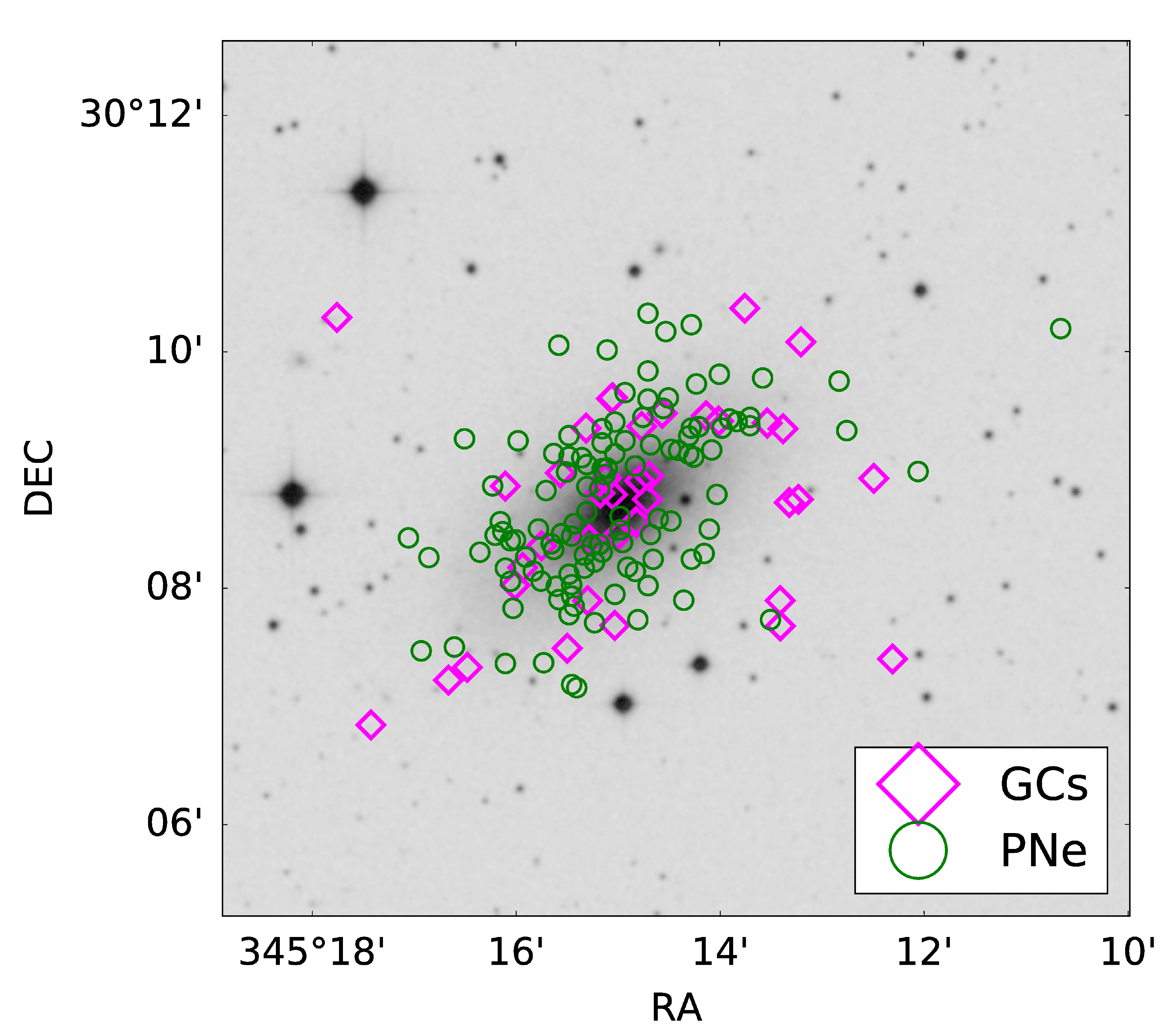} }}
    \caption{Spectroscopic tracers plotted over K-band images, from 2MASS, of the galaxies in this sample. Magenta diamonds are GCs from \citet{pota13} and green circles are PNe from the PN.S catalogue \citep{pns}.}%
    \label{fig:alltracers}%
\end{figure*}

In this work, we use spectroscopic and photometric data from GCs and PNe to model the kinematics of a sample of three nearby S0 galaxies. We have selected three non-cluster galaxies so that we could probe their formation mechanisms independently of any process commonly found in high-density environments (e.g. ram pressure stripping). In Table \ref{tab:data} we present the size of the globular cluster and planetary nebula samples, for each galaxy. The spatial distribution of the spectroscopic data is shown on Fig. \ref{fig:alltracers}. Additionally, in this section we describe the sources of the data and briefly discuss the galaxies' basic properties.

\subsection{Globular Clusters}

\label{gcdata}

The GC photometric and spectroscopic data analysed in this work comes from The SAGES Legacy Unifying Globulars and GalaxieS (SLUGGS) Survey\footnote{http://sluggs.swin.edu.au/Start.html} \citep{brodie14,forbes17}. SLUGGS is a wide-field spectroscopic and photometric survey of early-type galaxies undertaken using mainly the Subaru/Suprime-Cam imager and the Keck/DEIMOS spectrograph from the Keck-II-Telescope \citep{faber03}. SLUGGS goals revolve around studying the outer regions of early-type galaxies, where stellar light is faint. The survey has deep \textit{gri} imaging and spectroscopy around 8500 \text{\normalfont\AA} of 25 nearby early-type galaxies.

The photometric data of all galaxies studied in this work have been described in \citet{pota13}. 
The data for NGC\,2768 is a combination of Suprime-Cam/Subaru in the \textit{$R_{C}iz$} filters and the Advanced Camera for Surveys (ACS) from the Hubble Space Telescope (HST). 

NGC\,3115 data are described in detail in \citet{arnold}.The photometric data for this galaxy was obtained with the Subaru/Suprime-Cam in the \textit{gri} filters.\footnote{Additional spectra for NGC\,3115 are from the LRIS instrument on the Keck-I-Telescope \citep{Oke10} and the IMACS instrument from the Magellan telescope \citep{dressler11}.}

The photometric GC data for NGC\,7457 included in SLUGGS comes from \citet{hargis11}, obtained from observations in the \textit{BVR} filter with the WIYN/MiniMo imager \citep{saha00}. \citet{hargis11} also obtained spectroscopy for a sample of 20 NGC\,7457 GCs, although in this work we use an updated and larger spectroscopic sample for this galaxy's GCs obtained with the Keck/DEIMOS spectrograph and published in \citet{forbes17}. This new sample has spectroscopy for 40 GCs, which doubles the previously studied spectroscopic sample.

\subsection{Planetary Nebulae}

The PNe data for the galaxies studied in this work were obtained with The Planetary Nebulae Spectrograph (PN.S, \citet{pns}). We also make use of the PNe kinematics obtained in \citet{c13b} for the galaxies of our sample. 
The PN.S is a dedicated instrument mounted at the William Herschel Telescope (WHT) in La Palma, Spain. The instrument detects PNe using a technique based on counter-dispersed imaging (CDI), which enables us to obtain velocities and positions for PNe at the same time.

\begin{table}
\centering
\caption{Sizes of GC and PN samples used in this work. In the second column we show the number of the GCs, for each galaxy, for which only photometry is available; in the third column we have the number of GCs for which spectroscopy data are available and in the last column the number of PNe.}
\begin{tabular}{|l|l|l|l|}
\hline
Galaxy & $N_{GC}$ Photom. & $N_{GC}$ Spec. & $N_{PNe}$ \\ \hline
NGC\,2768 & 978 & 106 & 315 \\ 
NGC\,3115 & 781 & 150 & 188 \\ 
NGC\,7457 & 536 & 40 & 112 \\ \hline
\end{tabular}
\label{tab:data}
\end{table}

\subsection{Galaxies}

General properties of the galaxies studied in this work are presented in Table \ref{table:galaxies}. In this section we further discuss some of their features found in previous works.

\begin{table*}
\centering
\caption{General properties of the galaxies published in \citet{alabi17}. From left to right, the columns are: Galaxy designation, distance, systemic velocity, central stellar velocity dispersion within 1 kpc, ellipticity, environment (F=Field, G=Group), galaxy morphology from \citet{brodie14}, average luminosity-weighted age of the stellar population within 1 $R_e$ from \citet{mcdermid15}, effective radius, stellar mass and bulge-to-total light ratio from \citet{c13b}.}
\resizebox{\textwidth}{!}{%
\begin{tabular}{|l|l|l|l|l|l|l|l|l|l|l|}
\hline
Galaxy & Dist. (Mpc) & $V_{sys}$ (km/s) & $\sigma_{1kpc}$ (km/s) & $\epsilon$ & env. & morph. & Age  (Gyr) & $R_e$ (kpc) & $log(M_{*})$ (M$\odot$) & B/T \\ \hline
NGC\,1023 $^{\dagger}$ & 11.1 & 602 & 183 & 0.63 & F & S0 & 12.3 & 2.58 & 10.99 & 0.53 \\
NGC\,2768 & 21.8 & 1353 & 206 & 0.57 & G & E/S0 & 12.3 & 6.37 & 11.21 & 0.71\\ 
NGC\,3115 & 9.4 & 663 & 248 & 0.66 & F & S0 & 9.0 & 1.66 & 10.93 & 0.74\\ 
NGC\,7457 & 12.9 & 844 & 74 & 0.47 & F & S0 & 3.8 & 2.13 & 10.13 & 0.30\\ \hline
\end{tabular}
}
\raggedright{$^{\dagger}$ GCs and PNe kinematic analysis for this galaxy was done in \citet{c16} with the same method used in this work.}
\label{table:galaxies}
\end{table*}

\subsubsection{NGC\,2768}

NGC\,2768 is a group galaxy classified as an E6 by \citet{vauc91} and as an S0 1/2 by \citet{san94}, located at a distance of about 22 Mpc from us \citep{tully13} in the direction of the constellation of Ursa Major. It is part of the small Lyon Group of Galaxies 167 \citep{garcia93}, and has traces of ionised gas and a dust lane along the minor axis \citep{kim98}. It is interesting to add that the ionised gas havs been found to have different kinematics than the stars in the inner regions of the galaxy \citep{fried94}. \citet{pota13} found a bimodal distribution in colour for the sample of 978 GCs used in this work with a separation at $(Rc - z) = 0.57$ mag, obtained with KMM \citep{ashman94}.  

The GC system of this galaxy has been previously studied by 
\citet{pota13}, who found rotation for the red GCs and negligible rotation for the blue sub-population. \citet{forbes12} focused on the red GCs and found that this sub-population follows the radial surface density profile of the galaxy light and is compatible with the kinematics of the bulge component.

\subsubsection{NGC\,3115}

NGC\,3115 is the closest S0 galaxy to the MW, with a distance of 9.4 Mpc \citep{catiello14} and shows the clearer GC colour bimodality of our sample \citep{brodie12, pota13, catiello14, arnold}. \citet{pota13} found a colour separation  of photometric GCs at $(g-i) = 0.91$ mag.  

This galaxy is located in the field \citep{brodie14} and displays many interesting morphological structures, such as faint remnant spiral structures proposed by \citet{norris14} and re-detected recently using VLT/MUSE spectroscopy by \citet{guerou}. It has two faint companion galaxies \citep{doyle05}.  

\subsubsection{NGC\,7457}

NGC\,7457 is a field S0 galaxy \citep{brodie14} with a distance of 12.9 Mpc \citep{alabi17}. In contrast with the other galaxies in this sample, it shows no signs of bimodality in its GC population \citep{hargis11, pota13}. Previous studies of this galaxy also proposed a counter-rotating galaxy core \citep{sil02} and a possible major merger origin \citep{hargis11}. Furthermore, it presents the smallest amount of GCs of all galaxies in the sample, with a total number of $\sim 210 \pm$ 30 GCs \citep{hargis11}.

\section{Methods and Analysis}
\label{sec:methods}

In most galaxies, GCs can be separated into two sub-populations based on optical colour and metallicity, namely a red metal-rich population and a blue and metal-poor one. Different scenarios were proposed to explain the evolution of such sub-populations within an evolving galaxy, such as blue GCs being commonly found in the halo \citep{forbes97}, while red GCs are likely associated with the bulge, since they formed in a later phase along with the spheroidal component or migrated towards the centre of the galaxy after the gas-rich phase of galaxy evolution \citep{shapiro10}. By studying the GC kinematics we can find signatures of their origins, if they show a disc or spheroid-like behaviour. To reach this goal we need to recover the probability for every GC to belong to the galaxy as modelled with PNe, and to a given galaxy component. In this section we will describe all the steps needed to retrieve these probabilities. 

\subsection{Spheroid-disc Decomposition and Photometric Probabilities} 

\label{sec:decomp}

\begin{figure*}%
    \centering
    \subfloat[NGC\,2768]{{\includegraphics[width=6.5cm]{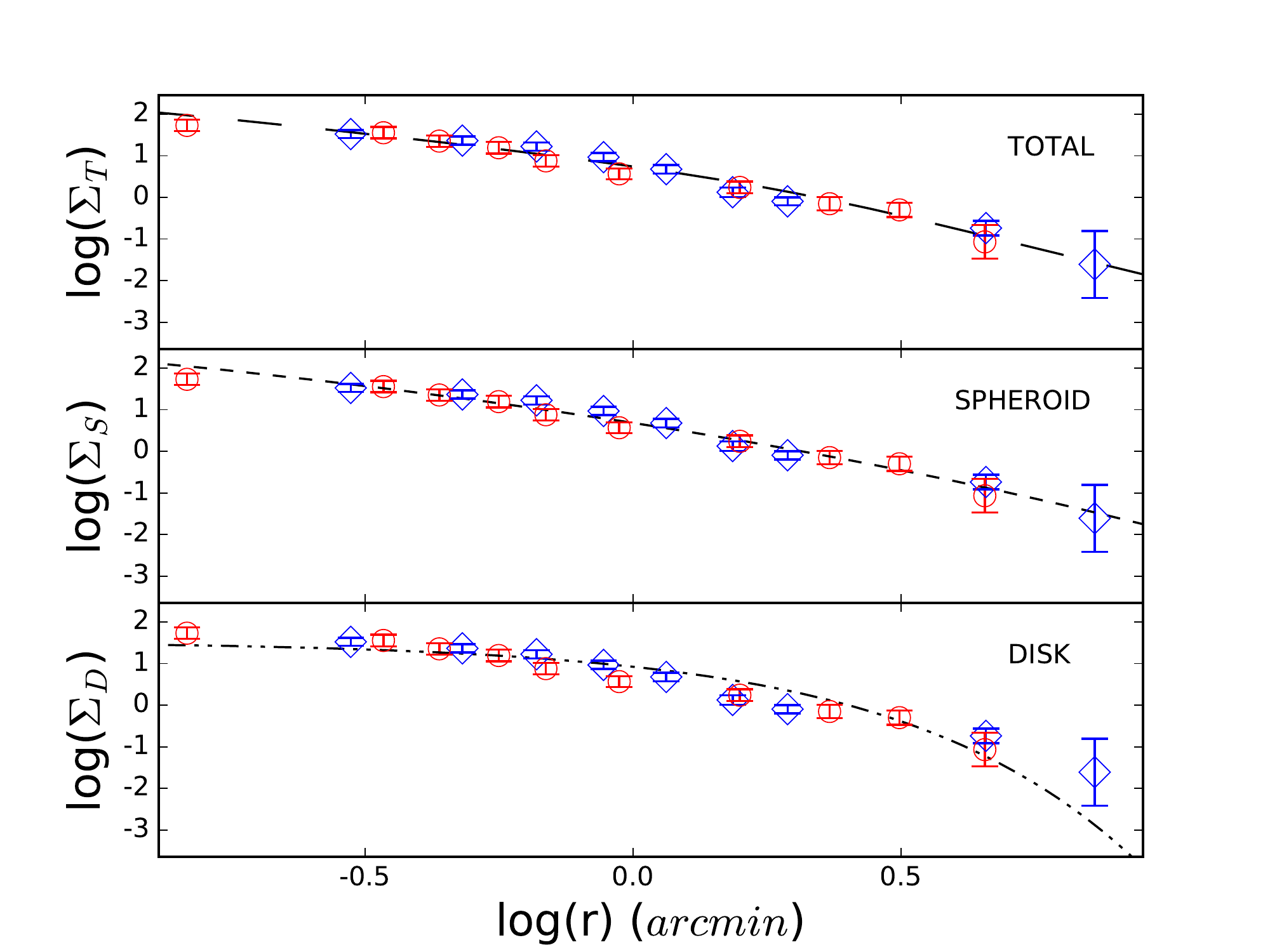}}}
    \subfloat[NGC\,3115]{{\includegraphics[width=6.5cm]{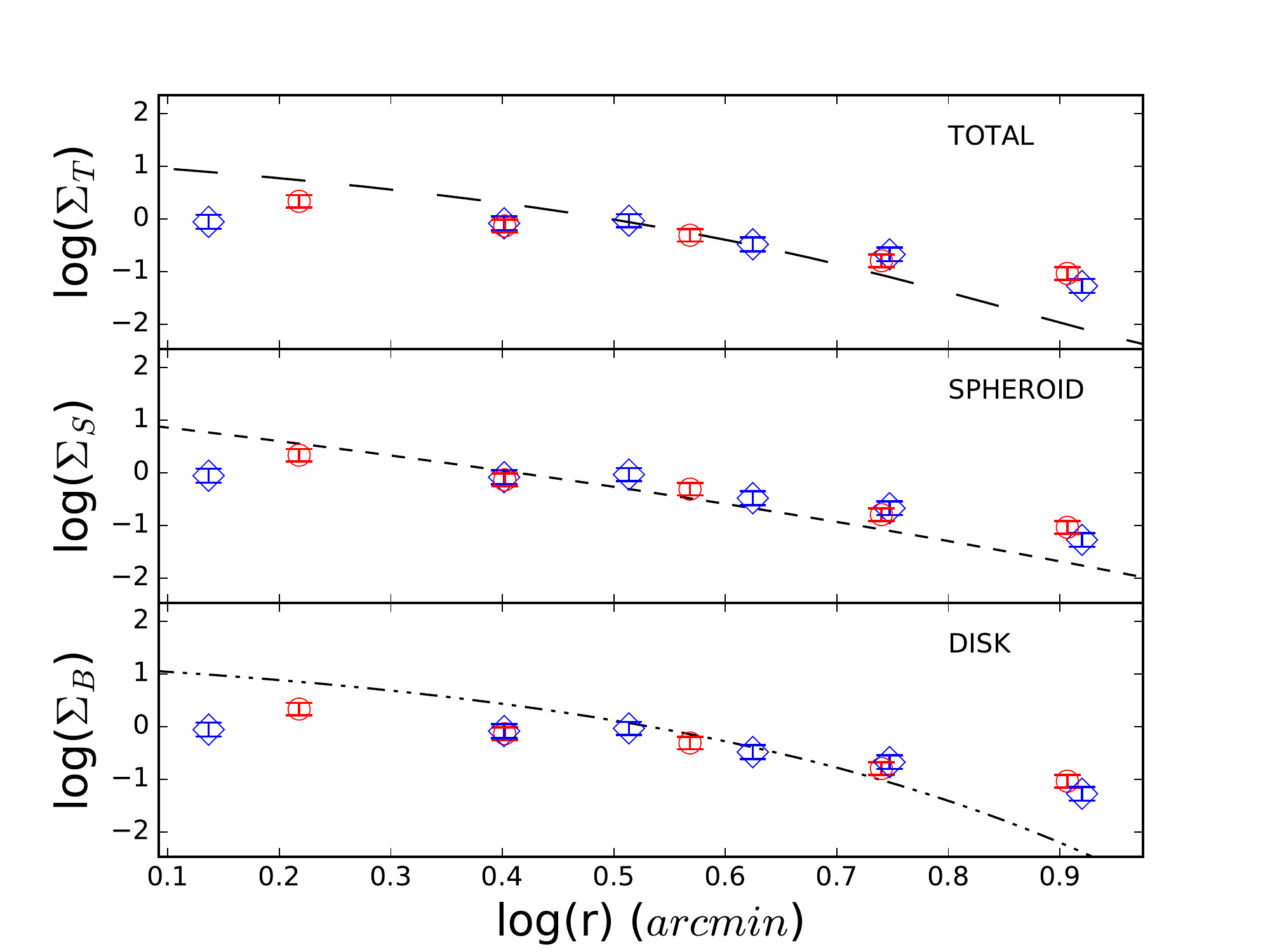}}}
    \subfloat[NGC\,7457]{{\includegraphics[width=6.5cm]{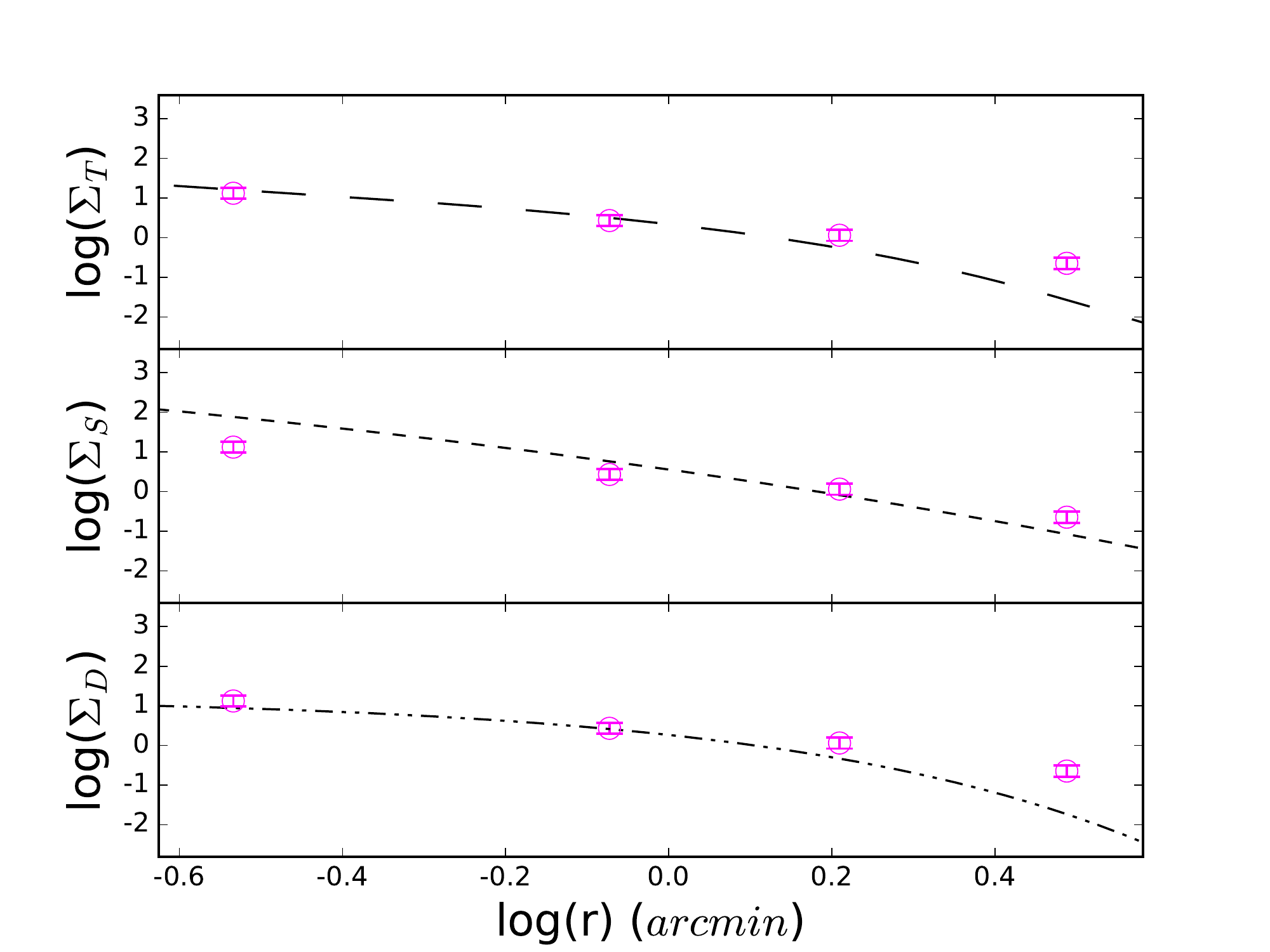}}}% 
    \caption{Photometric GC number density profiles (open diamonds) and galaxies' surface brightness profiles (following the \textsc{GALFIT} fit from \citet*{c13b}). Top panels compare photometric GC number density profiles with the total light profiles, while centre and lower panels compare the GC number density with spheroid and disc light profiles, respectively. For NGC\,2768 and NGC\,3115, both blue and red GC populations are shown, represented by red circles and blue diamonds, respectively, while for NGC\,7457 only the total population of GCs is shown, represented by magenta open diamonds. The light profile curves have been arbitrarily shifted along the y-axis to better match the GC number density in each panel. There is a slightly better comparison between GCs and spheroid profiles for all galaxies, although disc profiles also show a good agreement with GC density except at large radii. There is no clear difference between red and blue GCs.}
    \label{fig:density}%
\end{figure*}

\citet*{c13b} performed a photometric decomposition of the light of the galaxies in our sample, using a model comprising a disc and a spheroidal component (i.e., bulge and halo) and K-band images from the 2MASS catalogue \citep{skru06}. Using these results, we compare the light profiles of these galaxies with the radial number density of their associated GCs, as shown in Fig. \ref{fig:density}. For NGC\,2768 and NGC\,7457, GC radial number densities follow reasonably well, within errors, the light profiles of the galaxies and of their components. In the case of NGC\,3115, there is, however, a less obvious compatibility between GC density and the spheroid radial light profile. While the inner discrepancy might be due to incompleteness, i.e. the GCs get lost against the galaxy light in the central regions, at large radii, such difference might arise from treating the halo and the bulge as a single component in the model. 

In general, a disc and a spheroidal component are a good approximation to model the light of S0 galaxies. It is to be noted, nevertheless, that some galaxies have slightly more complex structures, and that the images used for the decomposition might be too faint to account for the halo component of the galaxy. Therefore, such a decomposition is prone to several systematic uncertainties and multiple approaches, such as the ones in \citet{savorgnan16} for NGC\,1023 and NGC\,3115. For the purpose of studying the kinematics of GC systems and PNe, the decomposition performed in \citet{c13b} is reasonable enough.

With the spheroid-disc light decomposition recovered in \citet*{c13b}, we also obtain preliminary photometric probabilities, $f_{i}$, for each tracer to belong to the spheroid model. We create an image dividing the spheroid by the total model of the galaxy light and we calculate the flux at the location of every GC in such image, within a circular aperture of radius equal to 3 pixels.
To obtain the final probabilities of a GC to belong to the spheroid or the disc we combine $f_{i}$ with the the GCs' kinematics, which will account for possible degeneracy or lack of precision of the photometric model \citep{c11} (see Section \ref{sec:chromo}).

\subsection{Kinematics}
\label{sec:like}
\begin{figure*}
	\centering
    \includegraphics[width=\textwidth]{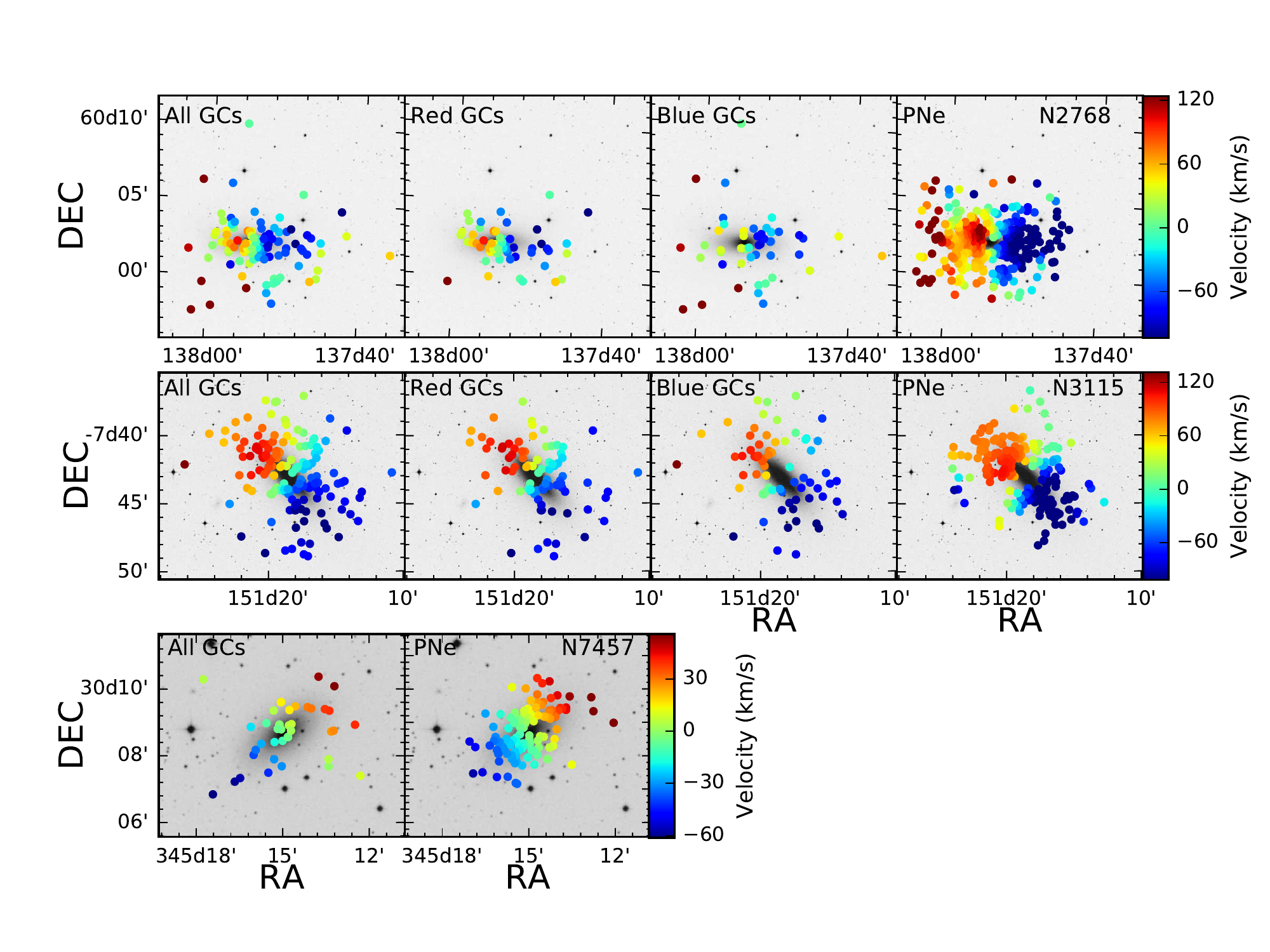}
    \caption{Smoothed velocity maps of sample galaxies PNe and GC populations, using Adaptive Kernel Smoothing \citep{coccato09}. For all galaxies rotation can be detected. However, most of the apparent rotation of the GC system of NGC\,2768 seems to be concentrated towards the centre and to be due to its red sub-population. NGC\,7457 has no signs of clear GC colour bimodality so the whole GC sample is shown.}%
    \label{fig:velocity}%
\end{figure*}

In Fig. \ref{fig:velocity} we show the smoothed velocity maps for the GC systems, obtained using Adaptive Kernel Smoothing (see \citet{coccato09} for details on using the technique). We detect clear rotation as measured from the GC systems for all galaxies. For NGC\,3115 this is true for both red and blue sub-populations, while we can notice that for the case of NGC\,2768 the rotation is supported by the inner red GCs.

To recover the kinematics of the GC and PNe systems of the galaxies in this work, we use Maximum Likelihood Estimation to find the best-fitting kinematic parameters, $\theta$, assuming a Gaussian velocity distribution for its GCs or PNe populations:

\begin{equation}\label{gaussian}
	F(v_{i};\theta)\propto exp\left[-\frac{(v_{i}-V_{los}(V)^2}{2\sigma^2}\right]
\end{equation}

where $v_{i}$ are the velocities of the discrete tracers, $V_{los}$ is the galaxy line-of-sight velocity, $V$ is the fitted rotation velocity and $\sigma$ is the fitted dispersion velocity.
We bin the data in elliptical annuli with the same ellipticity of the disc component for each galaxy (see Table\,\ref{table:galaxies}), and with approximately the same number of objects in every bin. In this way, we recover the kinematic profile with radius. Moreover, we iterate the MLE fit until all objects are within the 2.3 $\sigma$ confidence interval, discarding outliers in each run. This ensures the robustness of the fit even when dealing with a possible significant number of outliers (for example, accreted GCs).  

\begin{figure}%
    \centering
    \subfloat[NGC\,2768]{{\includegraphics[width=8cm]{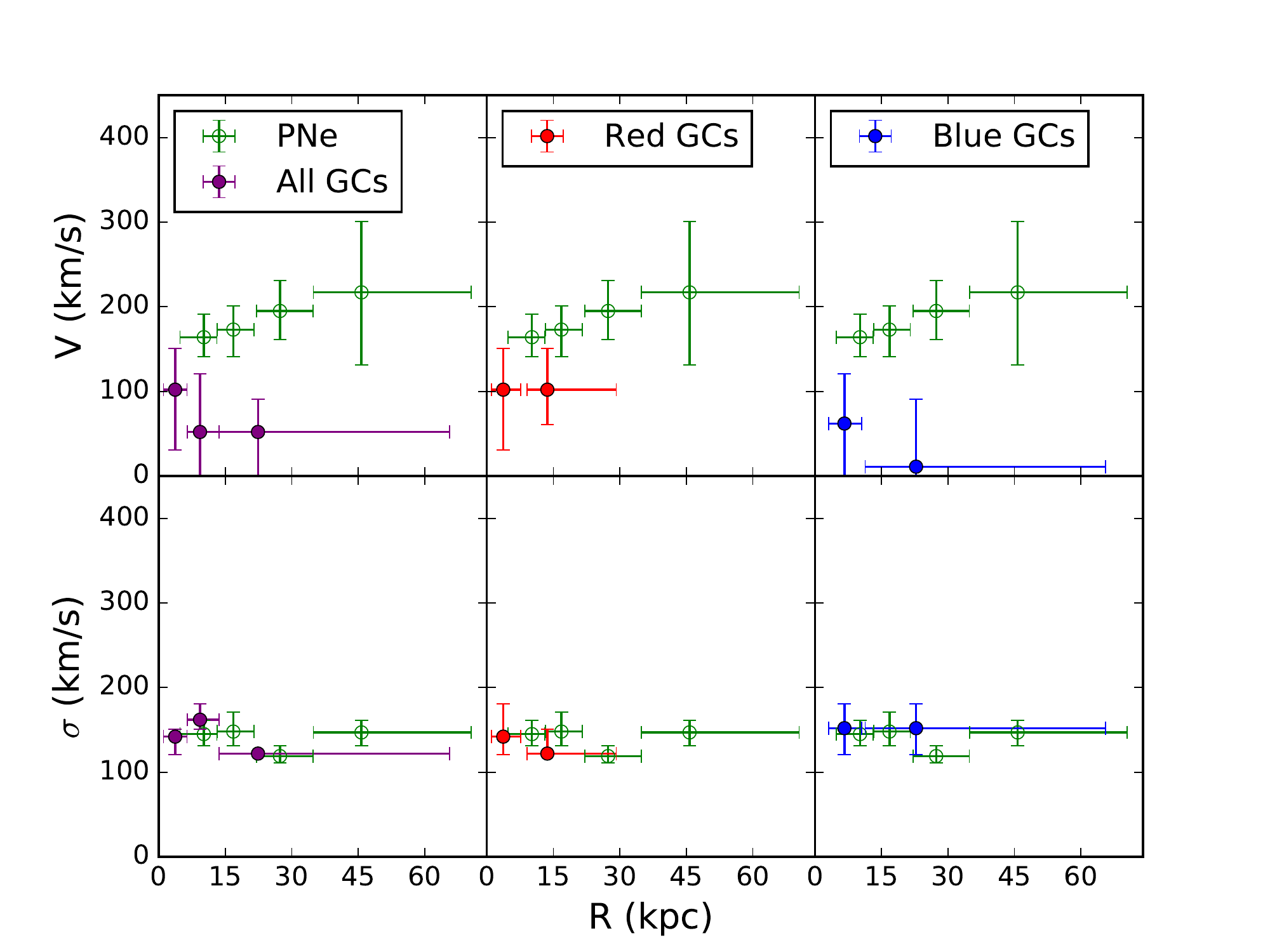} }} \\
    \subfloat[NGC\,3115]{{\includegraphics[width=8cm]{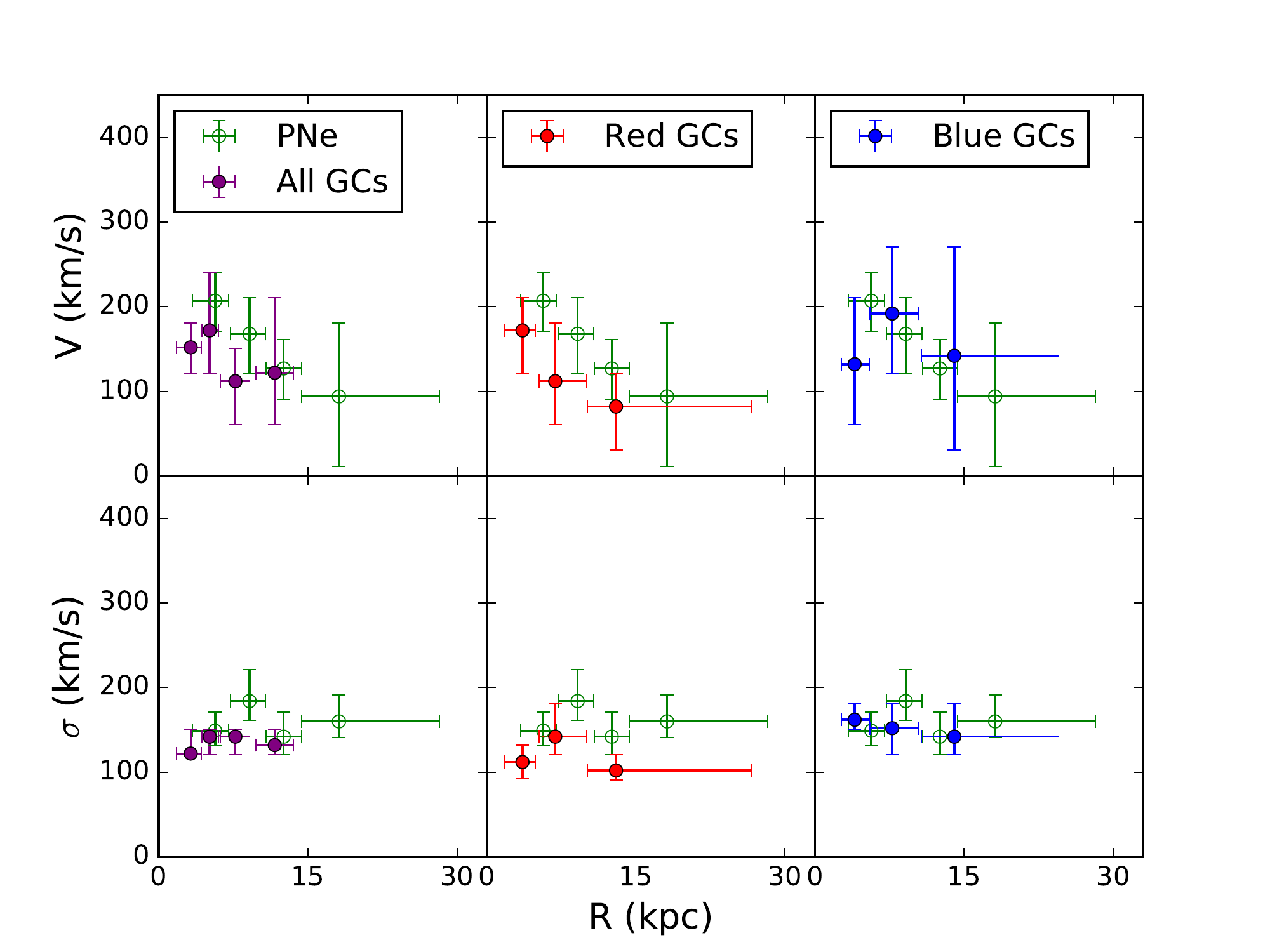} }} \\
    \subfloat[NGC\,7457]{{\includegraphics[width=8cm]{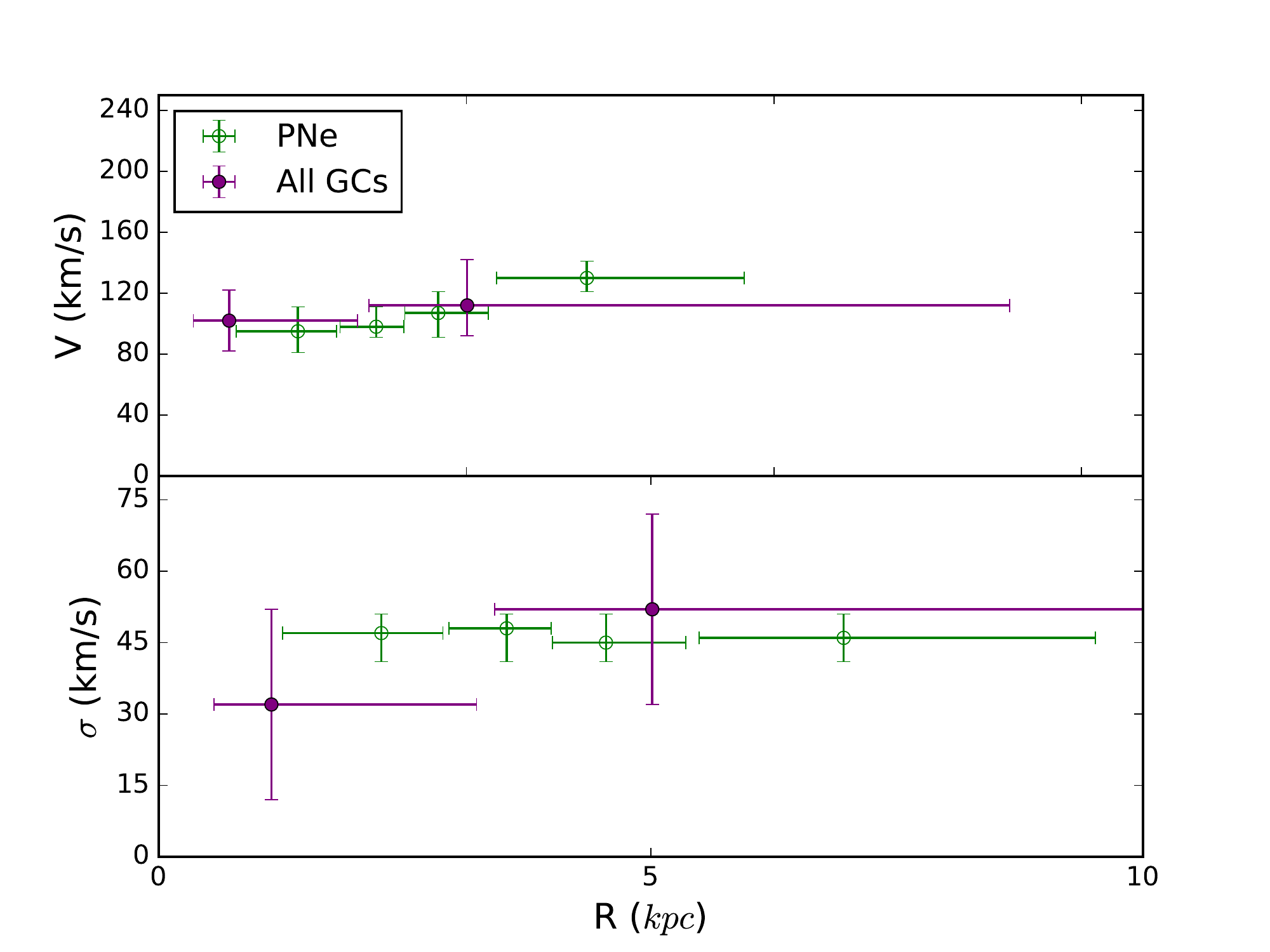} }}\\
    \caption{Rotation profiles and velocity dispersion for GCs and PNe obtained through the likelihood analysis, considering a one-component kinematic model. Vertical error bars are uncertainties and horizontal error bars represent bin-sizes. The data were divided in elliptical bins along circularised radius with approximately the same number of objects in each bin. The number of bins for each GC sub-population was designed aiming to ensure enough objects in each bin for the likelihood estimation. Note the distinct behaviour between the blue and red GC colour sub-populations of NGC\,2768 and the diverse trends for the kinematics of the different GC systems.}%
    \label{fig:like}%
\end{figure}

In Fig. \ref{fig:like} we present the results of the likelihood analysis for GCs and PNe. As far as the rotational velocity is concerned, the  galaxies have very distinct profiles: NGC\,2768 has strong kinematic discrepancies between its GC colour sub-populations, with red GCs showing rotation around 100 km/s and blue GCs showing negligible rotation. The rotational velocity of the GCs in NGC\,3115 decreases with radius consistently and independently of colour, as found for NGC\,1023 in \citet{c16}. For this galaxy, we also note that the rotational velocities of the PNe and GCs are consistent, despite the aforementioned model limitations. Lastly, the GCs of NGC\,7457 show a flatter rotational profile for both GCs and PNe. For the velocity dispersion, all galaxies show good agreement between PNe and GCs. This is expected since they share the same gravitational potential.

\subsection{Final probabilities using photometry and PNe kinematics}
\label{sec:chromo}
In this section we summarise the method introduced by \citet{c16}, and refer the reader to such work for further details.
\begin{figure*}%
	\centering
    \subfloat[NGC\,2768 Photometry]{{\includegraphics[width=7cm]{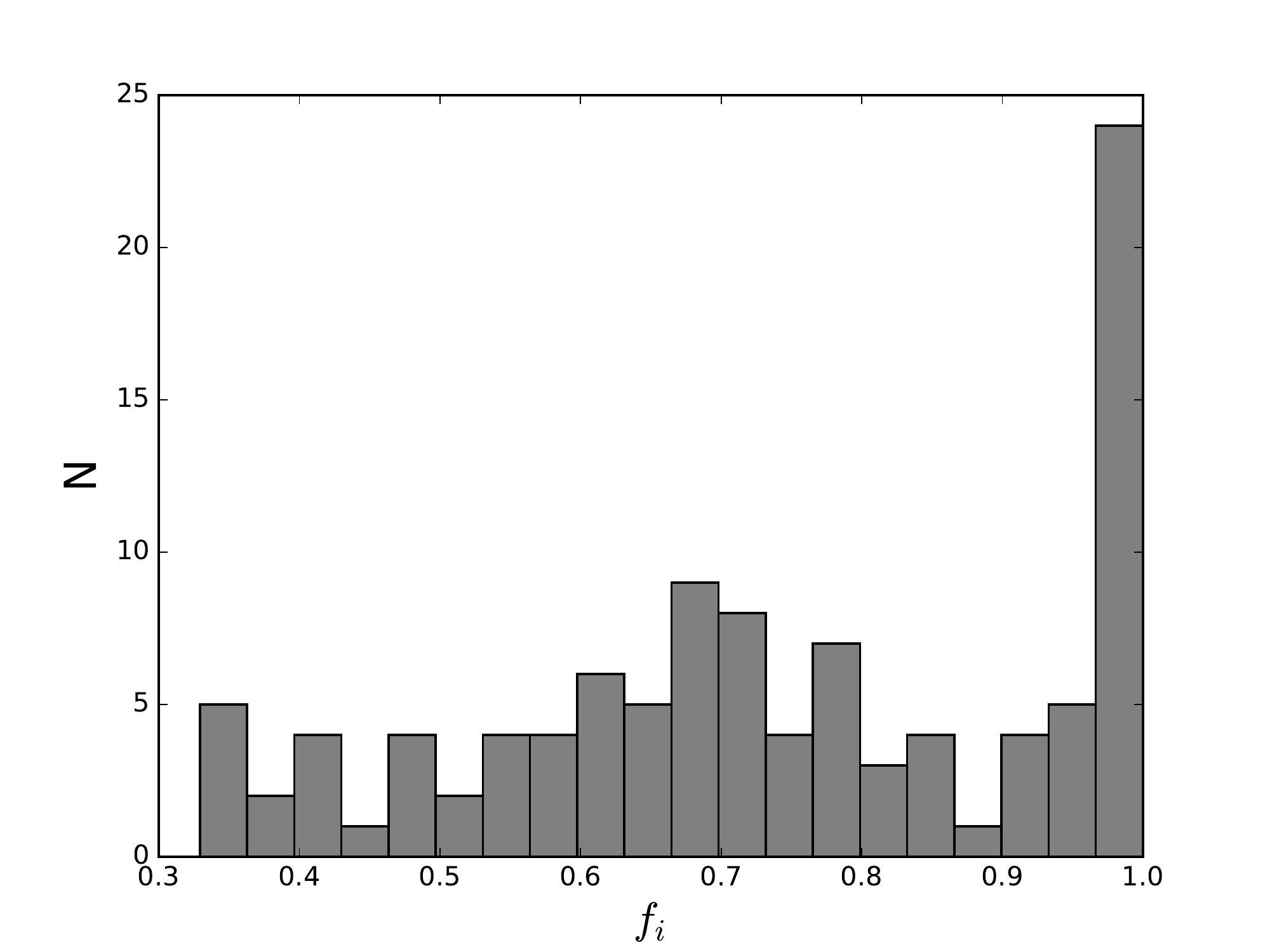}}}%
    \subfloat[NGC\,2768 Photometry + Kinematics]{{\includegraphics[width=7cm]{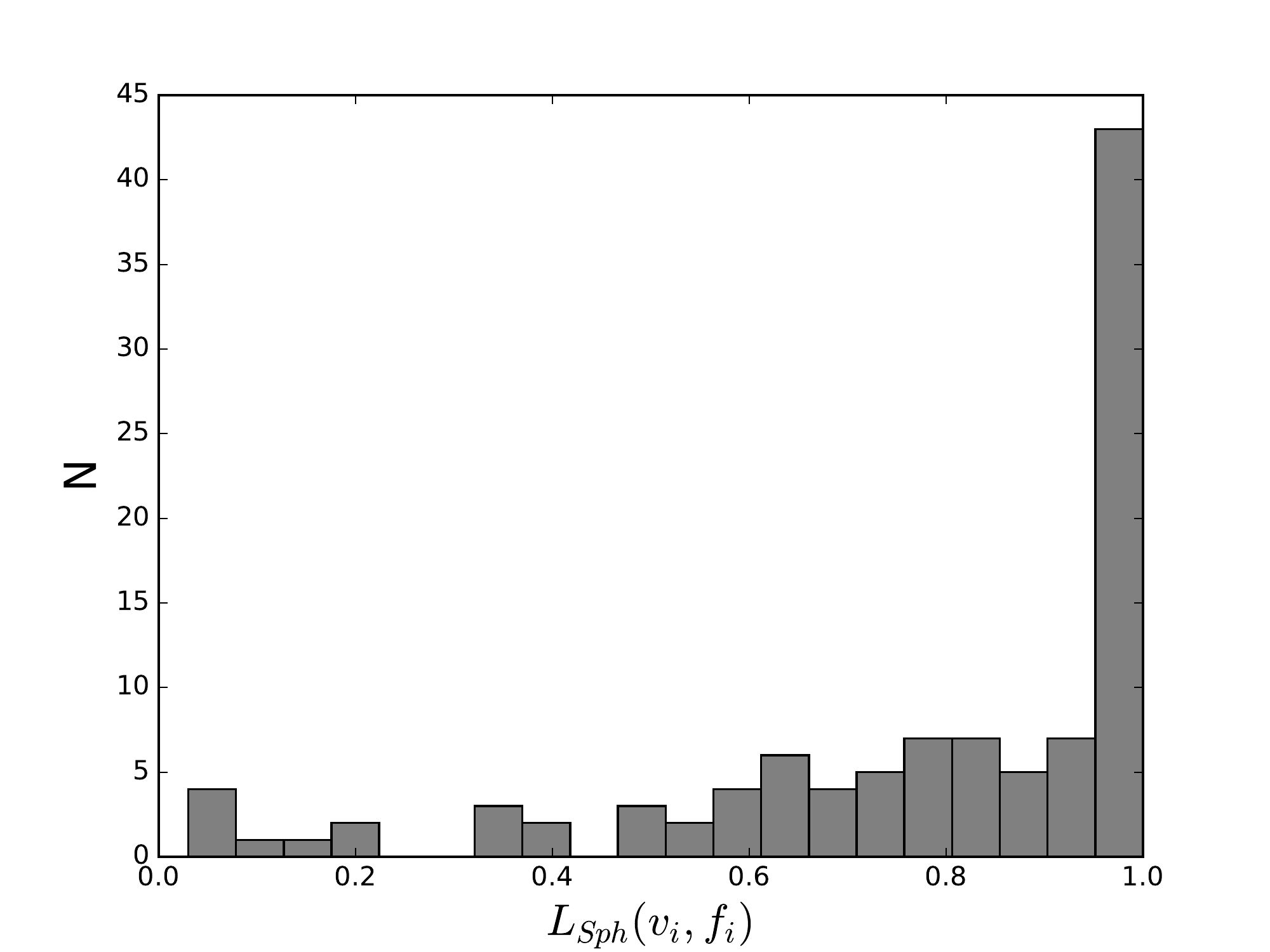}}}\\%
    \subfloat[NGC\,3115 Photometry Only]{{\includegraphics[width=7cm]{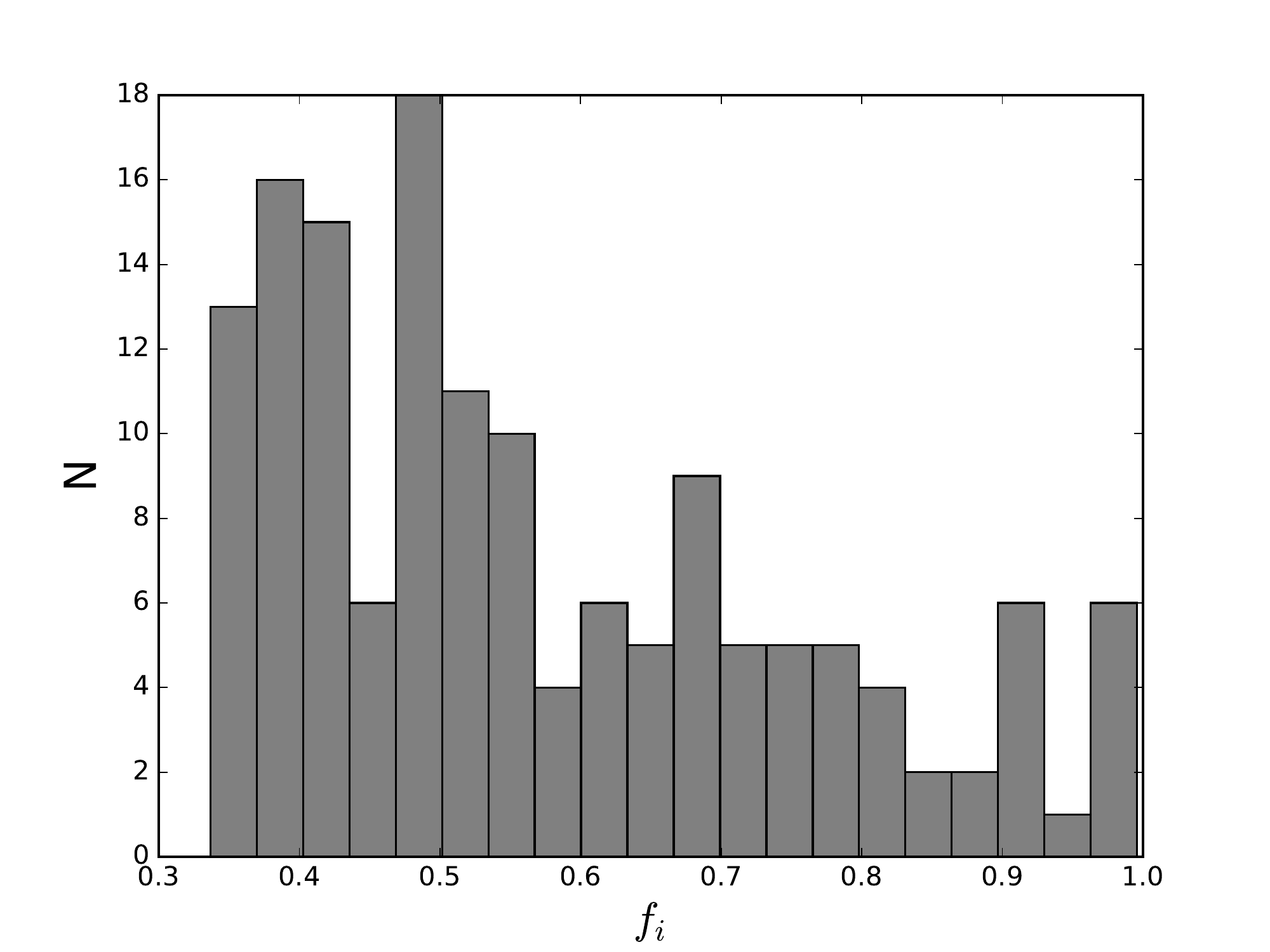}}}%
    \subfloat[NGC\,3115 Photometry + Kinematics]{{\includegraphics[width=7cm]{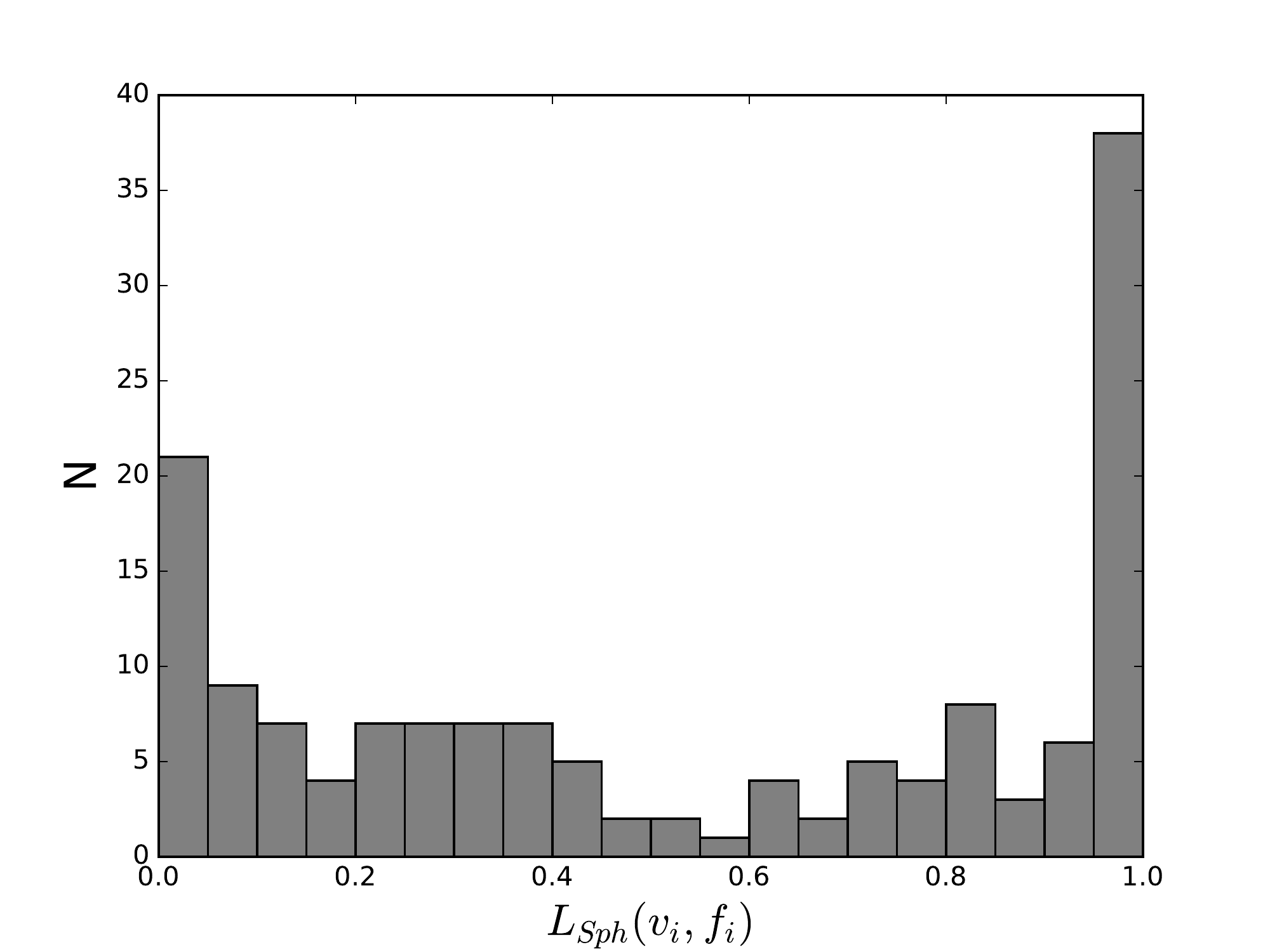}}}\\%
 	\subfloat[NGC\,7457 Photometry Only]{{\includegraphics[width=7cm]{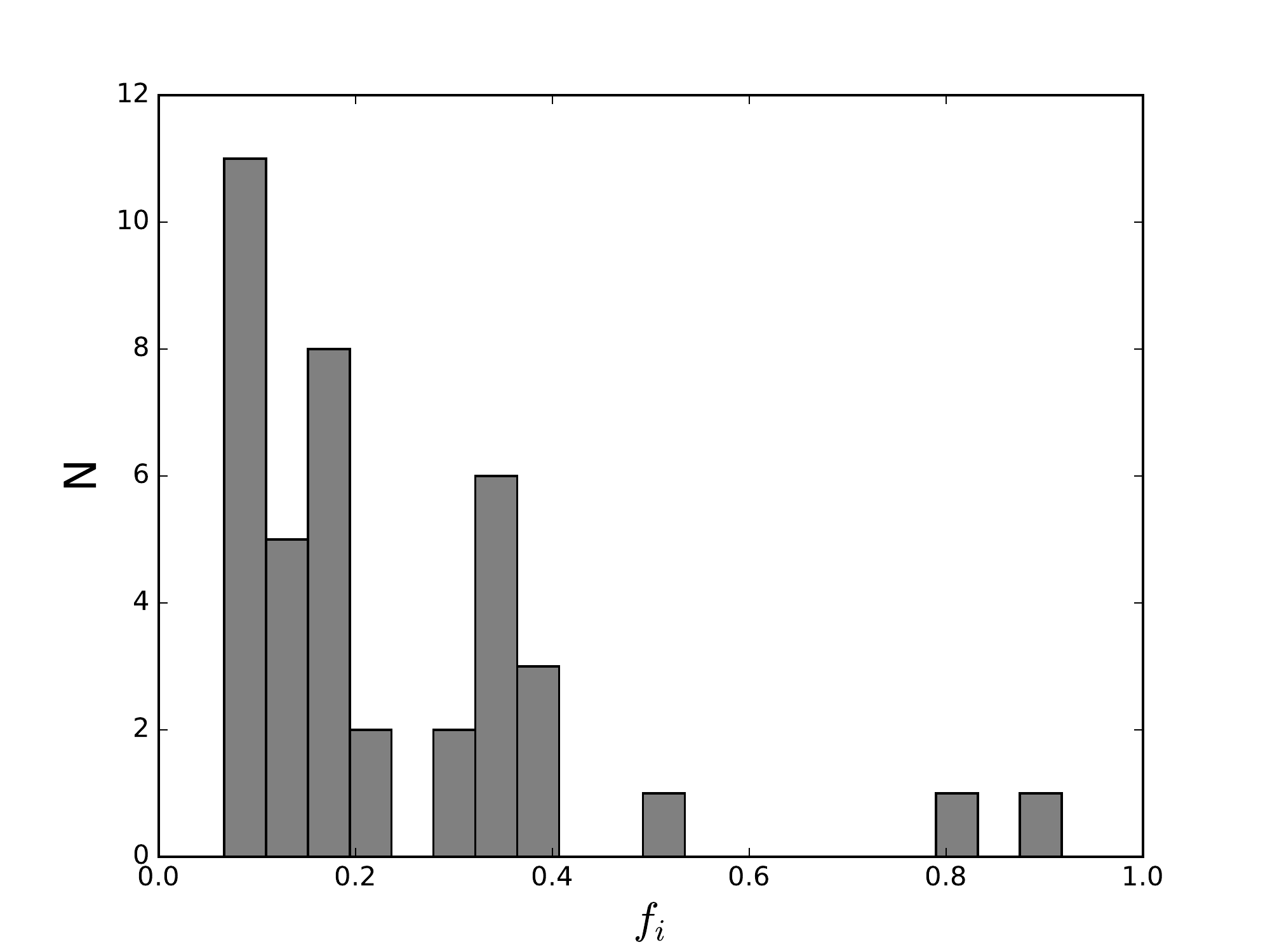}}}%
 	\subfloat[NGC\,7457 Photometry + Kinematics]{{\includegraphics[width=7cm]{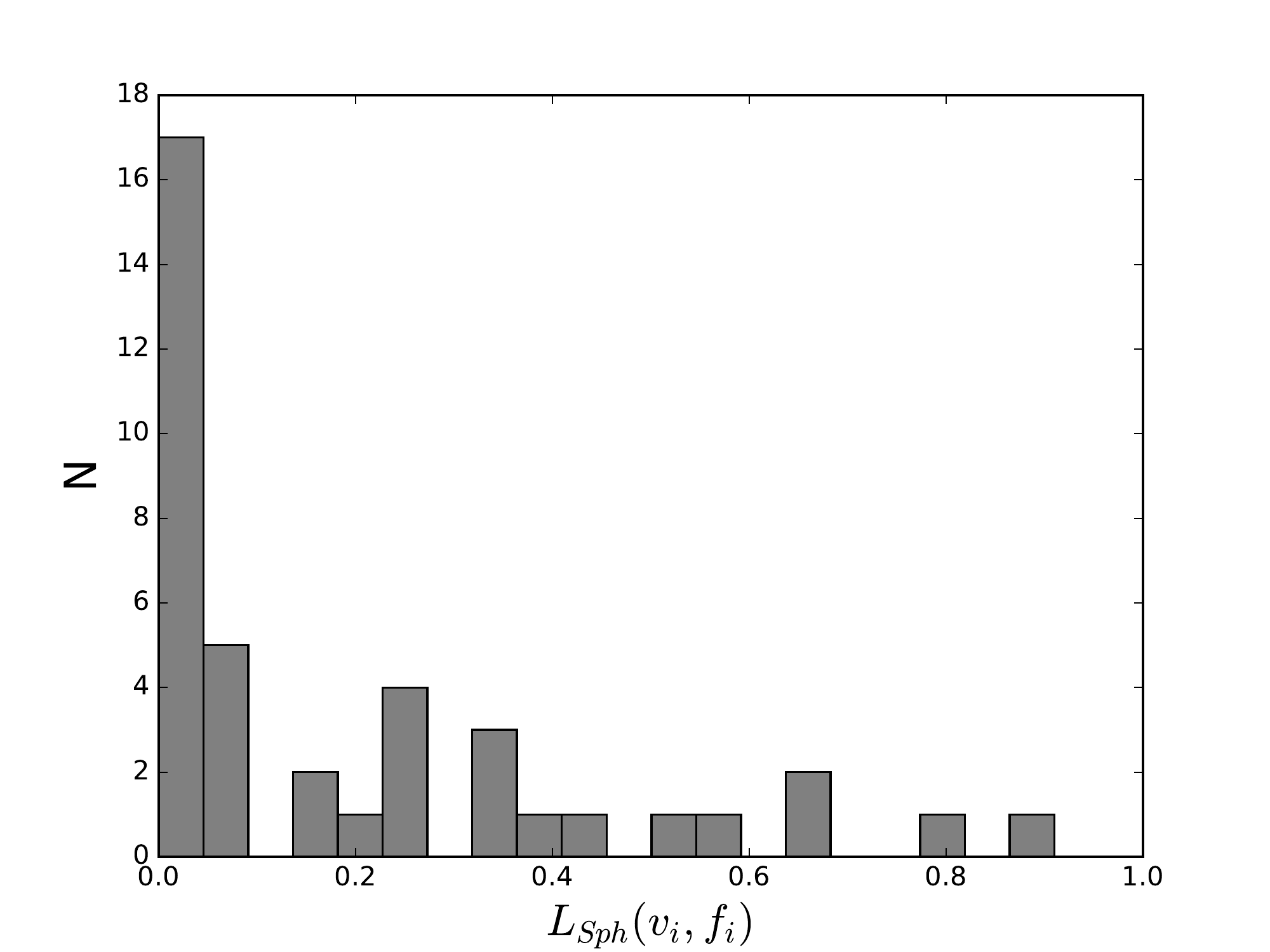}}}\\%
    \caption{The probability of GCs belonging to the host galaxy spheroid. \textit{Left panels}: probabilities from photometry, $f_i$; \textit{right panels}: probabilities from photometry and kinematics, $L_{sph}(v_i, f_i)$. Notice the decrease of objects with probability values around $0.5$ in the right panels compared to the left ones, showcasing the improvement of adding kinematic information to the photometric analysis. Also, notice the diverse cases presented in the right-side panels. Most of NGC\,2768 GCs are likely to belong to the spheroid, while NGC\,3115 has a significant number of GCs likely to be in the disc, as well as in the spheroid, and finally NGC\,7457 shows a dominant disc-like population of GCs.}%
    \label{fig:finalhist}%
\end{figure*}%

In \citet*{c13b}, the kinematics of the bulge and the disc of the galaxies in this work were modelled using a MLE fitting of the PNe velocity field. Using such kinematic models we can derive the probabilities that every GC belongs to the galaxy disc or spheroid, solving the following equation:

\begin{multline}\label{lpne}
    \mathcal{L}(v_{i};V, \sigma_{r}, \sigma_{\phi}, \sigma_{sph})\propto\frac{f_{i}}{\sigma_{sph}}exp\left[-\frac{v_{i}^2}{2\sigma_{sph}^2}\right]+ \\
    +\frac{1-f_{i}}{\sigma_{los}(\sigma_r, \sigma_{\phi})}exp\left[-\frac{(v_{i}-V_{los}(V))^2}{2\sigma_{los}(\sigma_r, \sigma_{\phi})^2}\right].
\end{multline}

where $v_i$ are the individual GC velocities, $V$ is the galaxy rotational velocity as derived from PNe, $V_{los}$ and $\sigma_{los}$ are the line-of-sight velocity and dispersion velocity, respectively.
The radial and azimuthal components of the velocity dispersion of the disc component are given by $\sigma_r$ and $\sigma_{\phi}$, while $\sigma_{sph}$ is the dispersion velocity for the spheroid component (also derived from PNe). The $f_i$ values are the probabilities obtained in Section \ref{sec:decomp} for every GC to belong to the galaxy spheroid component based on photometry only.

In Fig. \ref{fig:finalhist}, we show the recovered probabilities of the GCs to belong to the spheroid, $\mathcal{L}_{sph}(v_i, f_i)$, obtained normalising the first term on the right side of equation \ref{lpne}, in comparison with the probabilities to belong to the spheroid retrieved from the photometry only, $f_{i}$. We notice a  decrease in probability values around 0.5 present in the histogram that show probabilities obtained from photometry only, in comparison with the probabilities obtained with photometry and kinematics. This shows the core improvement of the MLE method over the photometric probability approach justifying our choice of a simple bulge-disc decomposition.  

We calculate the total number of GCs that belong to the galaxy disc or spheroid by summing-up the probabilities $L_{sph}(v_i, f_i)$ and the total number of disc GCs is obtained summing (1-$L_{sph}(v_i, f_i)$). These values are outlined in Table \ref{tab:discspheroid}, for the total, red and blue sub-populations.

Moreover, the likelihood fit also offers us the advantage of detecting objects that are outliers, i.e., not likely to be compatible with the host galaxy kinematics, as traced by PNe in \citet*{c13b}. We set the likelihood threshold at 2.3$\sigma$ (see \citet{c16}). Within the entire GC spectroscopic sample, such objects account for 10 GCs in NGC\,2768, 32 GCs in NGC\,3115 and 4 GCs in NGC\,7457. This means that 21.5\% of NGC\,3115 GCs are not likely to be compatible with the galaxy model (previously obtained with PNe and the K-band image). Therefore we can explain such objects in two ways. One possibility is that they might have been recently accreted. Another option is that they might just belong to the galaxy halo, which is not separately included in this study, but is treated together with the bulge as a single spheroidal component. Nevertheless, this number of rejected objects in NGC\,3115 is a higher amount than the 9.4\% rejected objects in NGC\,2768 and 10\% in NGC\,7457.

\section{Results }

\label{sec:results}

\subsection{Colour and Kinematics}

\begin{figure}%
	\centering
    \subfloat[NGC\,2768]{{\includegraphics[width=0.5\textwidth]{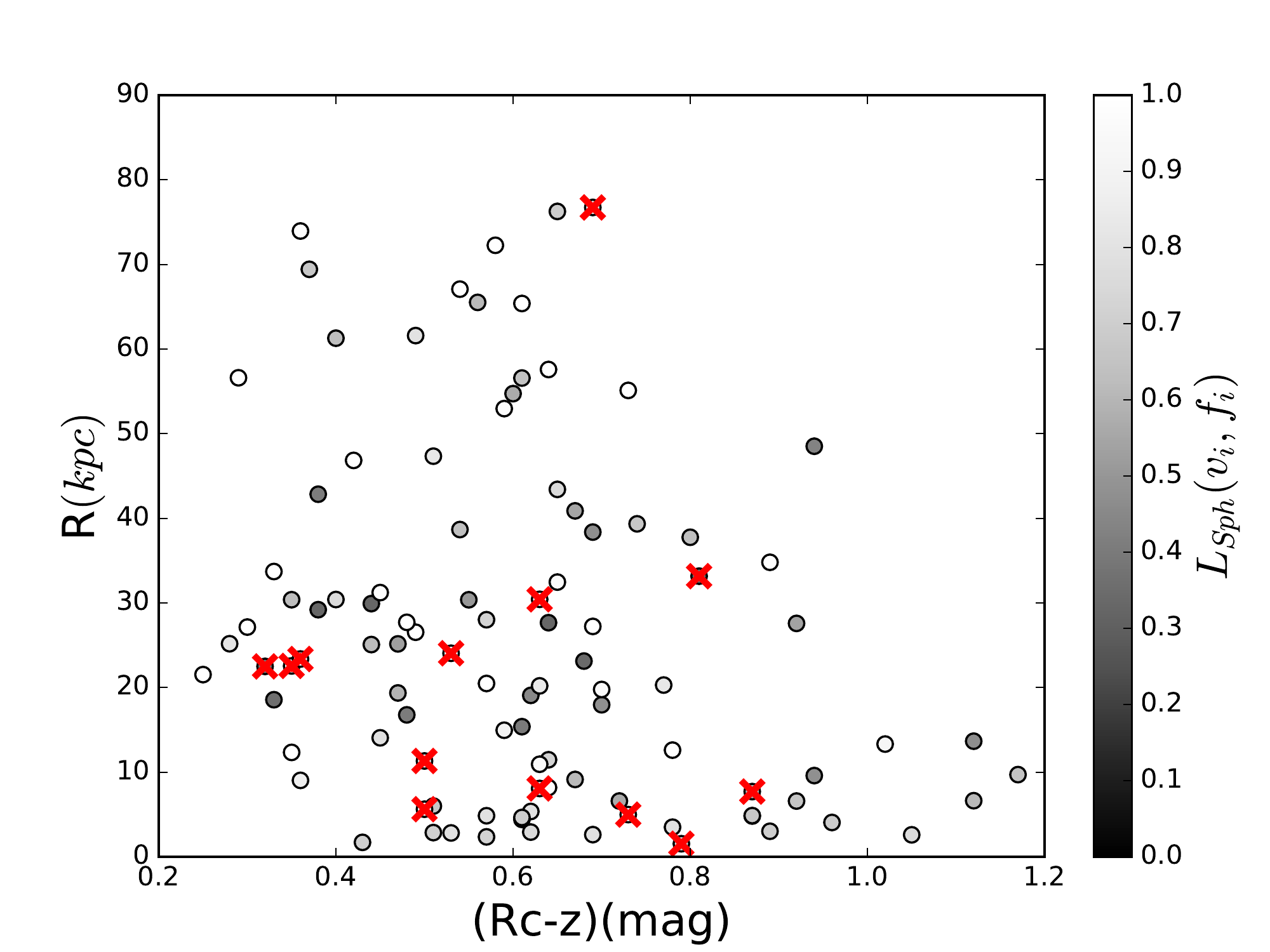}}}\\%
    \subfloat[NGC\,3115]{{\includegraphics[width=0.5\textwidth]{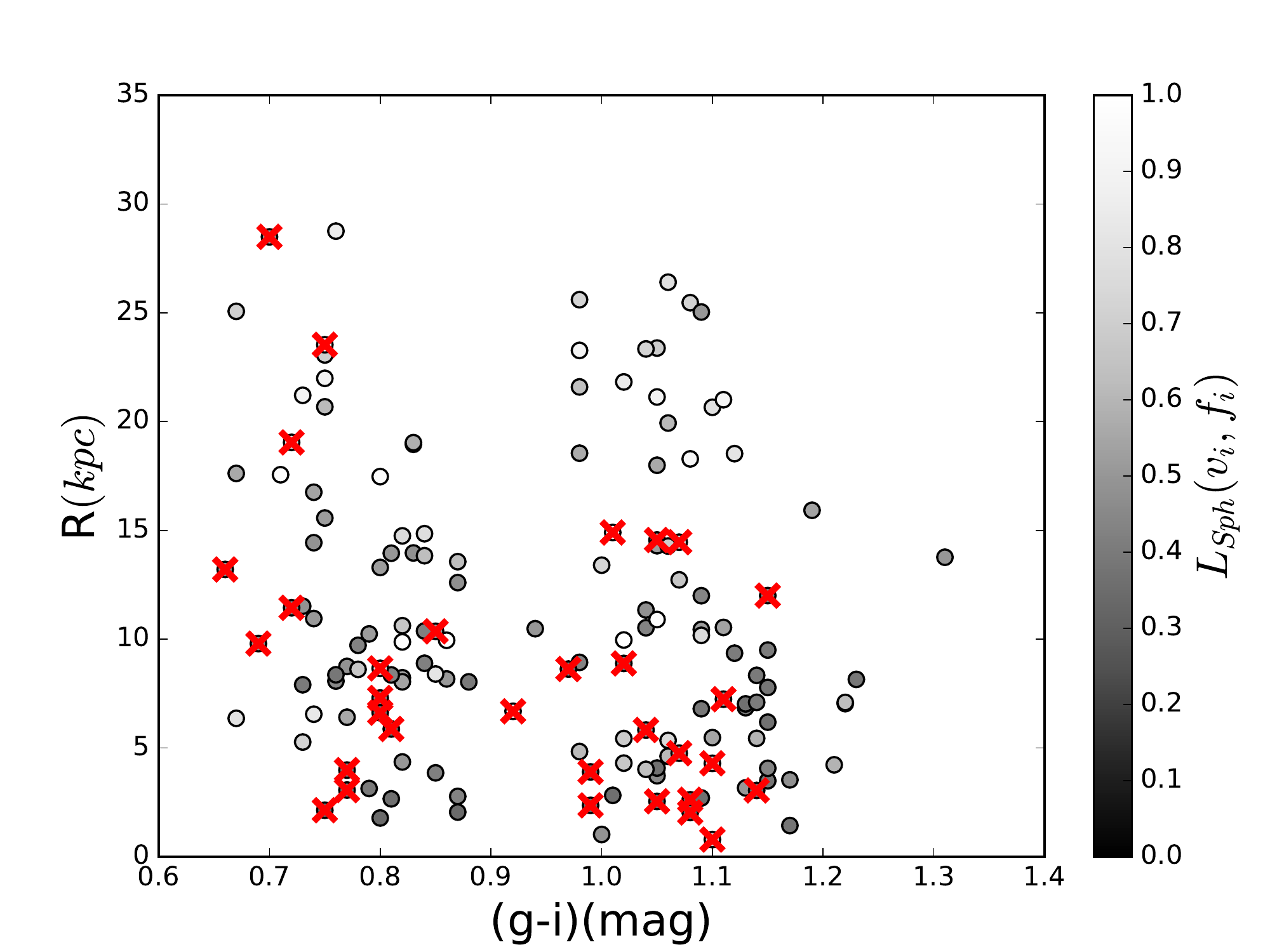}}}\\%
 	\subfloat[NGC\,7457]{{\includegraphics[width=0.5\textwidth]{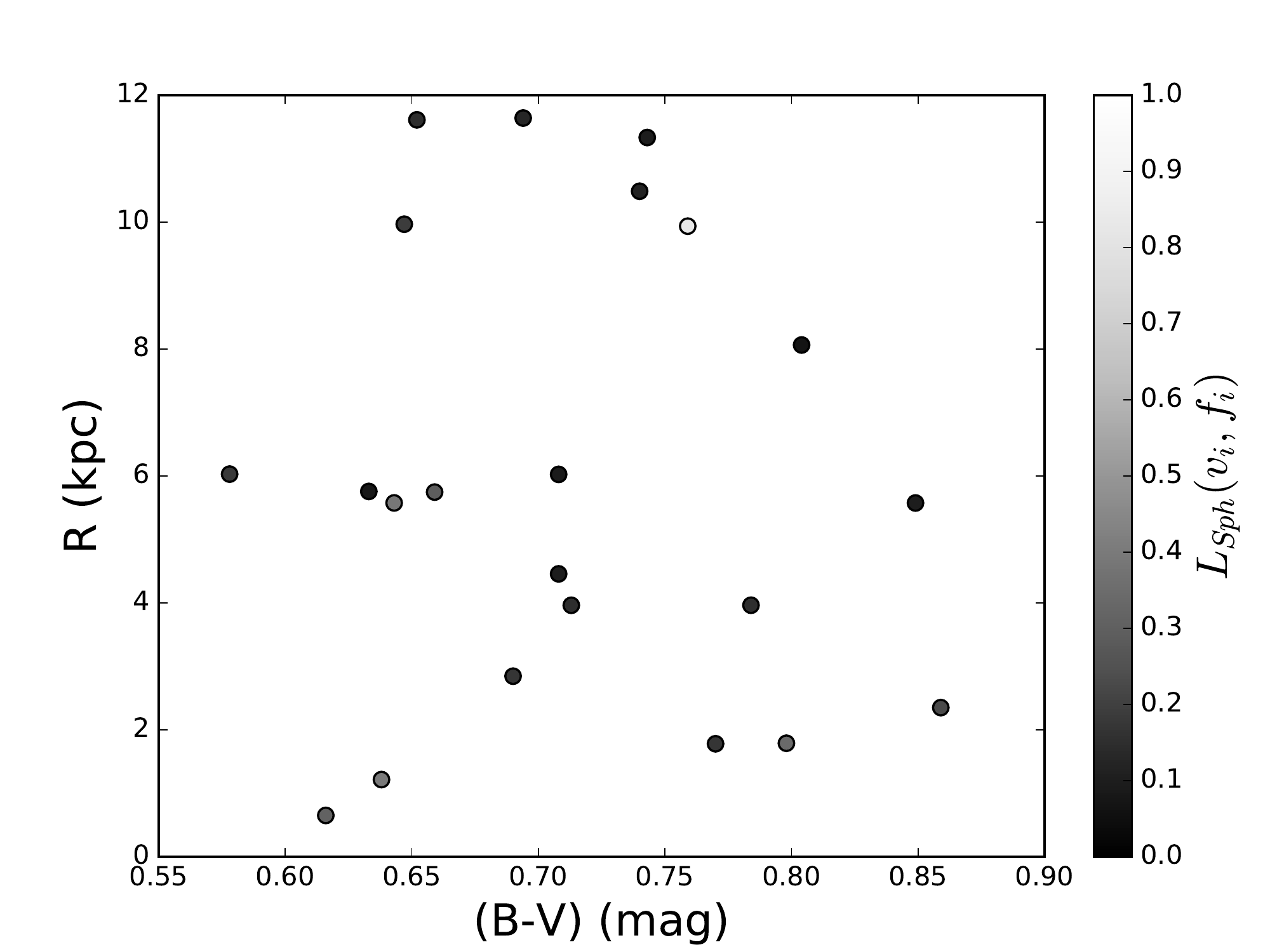}}}\\%
    \caption{Radial distribution of GCs, colour coded by $L_{Sph}(v_{i}, f_{i})$, the probability of belonging to the spheroid. Red X-shaped markers represent rejected objects with kinematics not compatible with their host galaxy kinematics following the likelihood analysis. For NGC\,7457 we lack colour information for some GCs. Therefore
    the rejected objects are not shown.}%
    \label{fig:rcolf}%
\end{figure}%

In Fig. \ref{fig:rcolf} we show the radial distribution of GCs versus their colours, colour coded according to their probability of belonging to the spheroid, $L_{Sph}(v_{i}, f_{i})$.  
 
Figure \ref{fig:rcolf} clearly shows that we are looking at three galaxies with very distinct GC systems. As mentioned in Section \ref{sec:data}, NGC\,3115 is clearly bimodal in colour, similarly to NGC\,2768, while NGC\,7457 shows no bimodality in colour for its GCs. Moreover, for NGC\,2768 17\% of GCs are associated with the disc component while NGC\,3115 has a similar number of disc-like and spheroid-like GCs, and NGC\,7457 has 15\% GCs associated with the spheroid component. We find no correlation between the colours of the GCs and their kinematics, since in NGC\,2768 both the populations of red and blue GCs have spheroid-like kinematics \footnote{See discussion in section \ref{sec:chromo} and Fig.\ref{fig:like} about differences between the red and blue GCs kinematics regarding the rotation profiles}. In NGC\,3115 there are blue and red GCs associated with both the spheroid and the disc. For NGC\,7457, the GC system does not appear bimodal neither in colour nor kinematics. For comparison, NGC\,1023 have 19 red and 18.6 blue GCs belonging to the disc, along 13 red and 25.4 blue GCs belonging to the spheroid \citep{c16}.
On the other hand, we notice that most of the GCs that are more likely to have disc-like kinematics are located in the inner regions of the galaxies, within 10 kpc. It is possible, however, to find some disc-like GCs at larger radii ($r>5 R_{e}$). For all galaxies, rejected GCs are scattered in the plot, suggesting no correlation among colour, kinematics and radius. Those objects will be discussed in more detail in the following section.  

\begin{table*}
\centering
\caption{Number of GCs associated with each of the host galaxy components, following the kinematic analysis described in Sec. \ref{sec:like}. GCs are also divided by colour sub-populations in the case of NGC\,2768 and NGC\,3115.}
\label{tab:discspheroid}
\begin{tabular}{|l|l|l|l|l|l|l|}
\hline
\multicolumn{1}{|c|}{\multirow{2}{*}{Galaxies}} & \multicolumn{3}{c|}{Disc}                               & \multicolumn{3}{c|}{Spheroid}                           \\ \cline{2-7} \\
\multicolumn{1}{|c|}{}                          & Red GCs                & Blue GCs               & Total & Red GCs                & Blue GCs               & Total \\ \hline
NGC\,2768                                        & 14.8                      & 10.0                     & 24.8    & 40.2                     & 27.9                     & 68.1    \\ \hline
NGC\,3115                                        & 26.8                     & 21.4                     & 48.2    & 37.2                     & 31.5                     & 68.7    \\ \hline
NGC\,7457                                        & \multicolumn{1}{c|}{-} & \multicolumn{1}{c|}{-} & 27.3    & \multicolumn{1}{c|}{-} & \multicolumn{1}{c|}{-} & 8.8    \\ \hline
\end{tabular}
\end{table*}

\subsection{1D phase-space diagrams}

\begin{figure}
	\centering
    \subfloat[NGC\,2768]{{\includegraphics[width=0.45\textwidth]{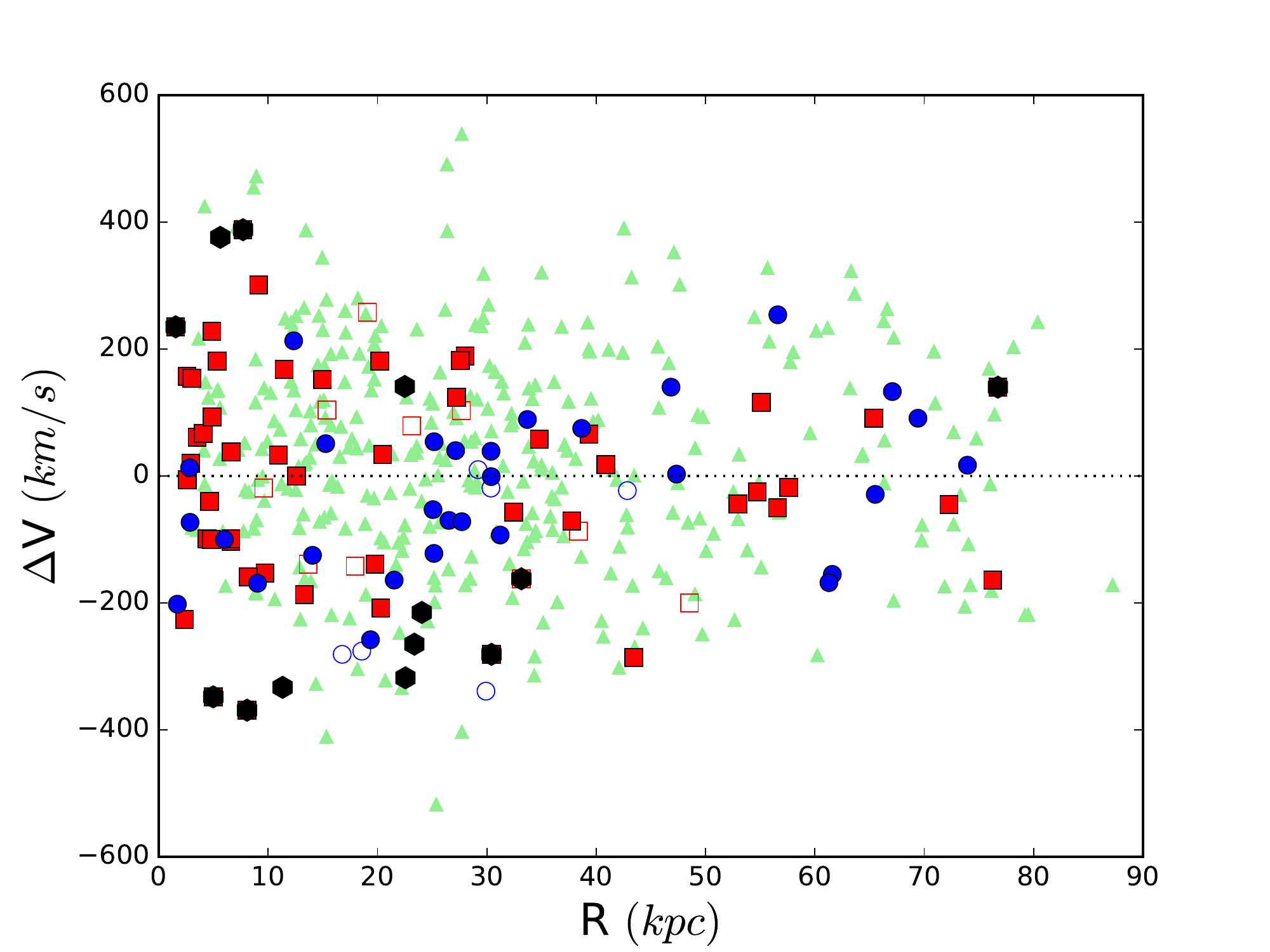}}}\\%
    \subfloat[NGC\,3115]{{\includegraphics[width=0.45\textwidth]{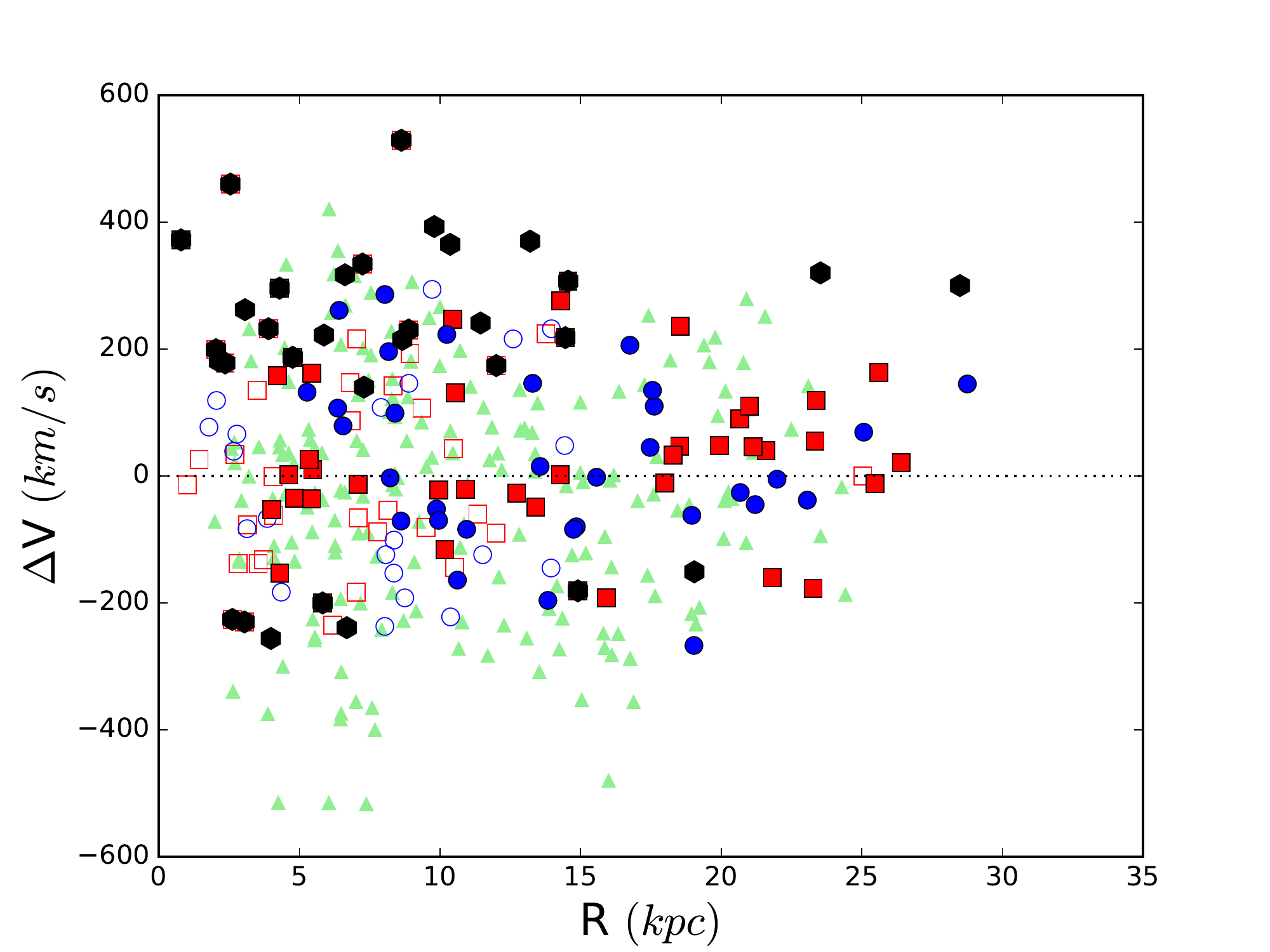}}}\\%
 	\subfloat[NGC\,7457]{{\includegraphics[width=0.45\textwidth]{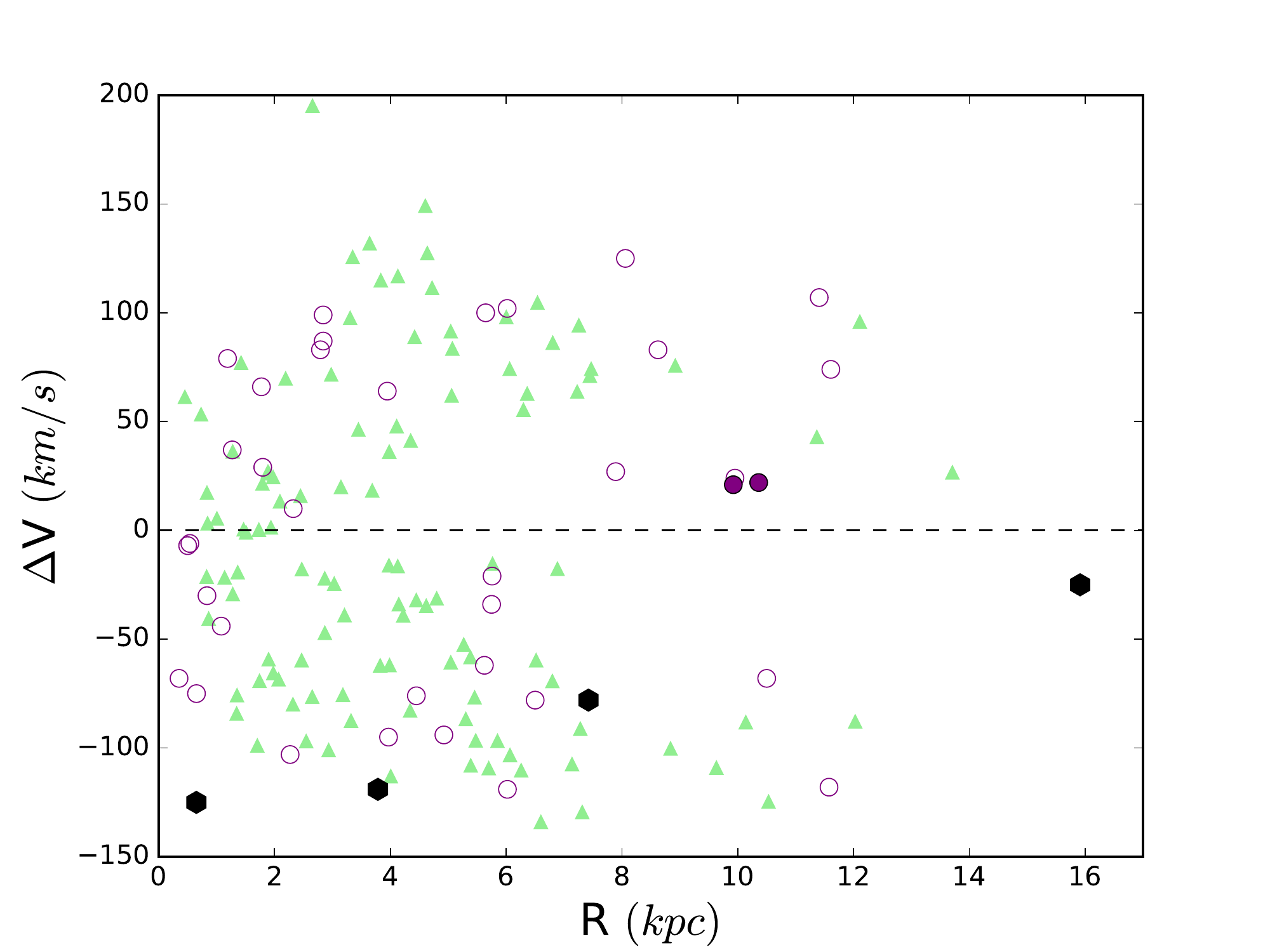}}}%
    \caption{1D phase-space diagrams for GCs and PNe. Blue and red GCs are represented by blue circles and red squares, respectively. GCs likely to belong to the spheroid are represented by filled markers, and GCs with high probability of being part of the disc are represented by open markers. Green filled triangles are PNe. For NGC\,7457, due to its lack of bimodality in colour, we show only disc and spheroid populations. Objects marked as black hexagons are rejections from the kinematic fit.}%
    \label{fig:phasevel}%
\end{figure}

\begin{figure}
	\centering
    \subfloat[NGC\,2768]{{\includegraphics[width=0.45\textwidth]{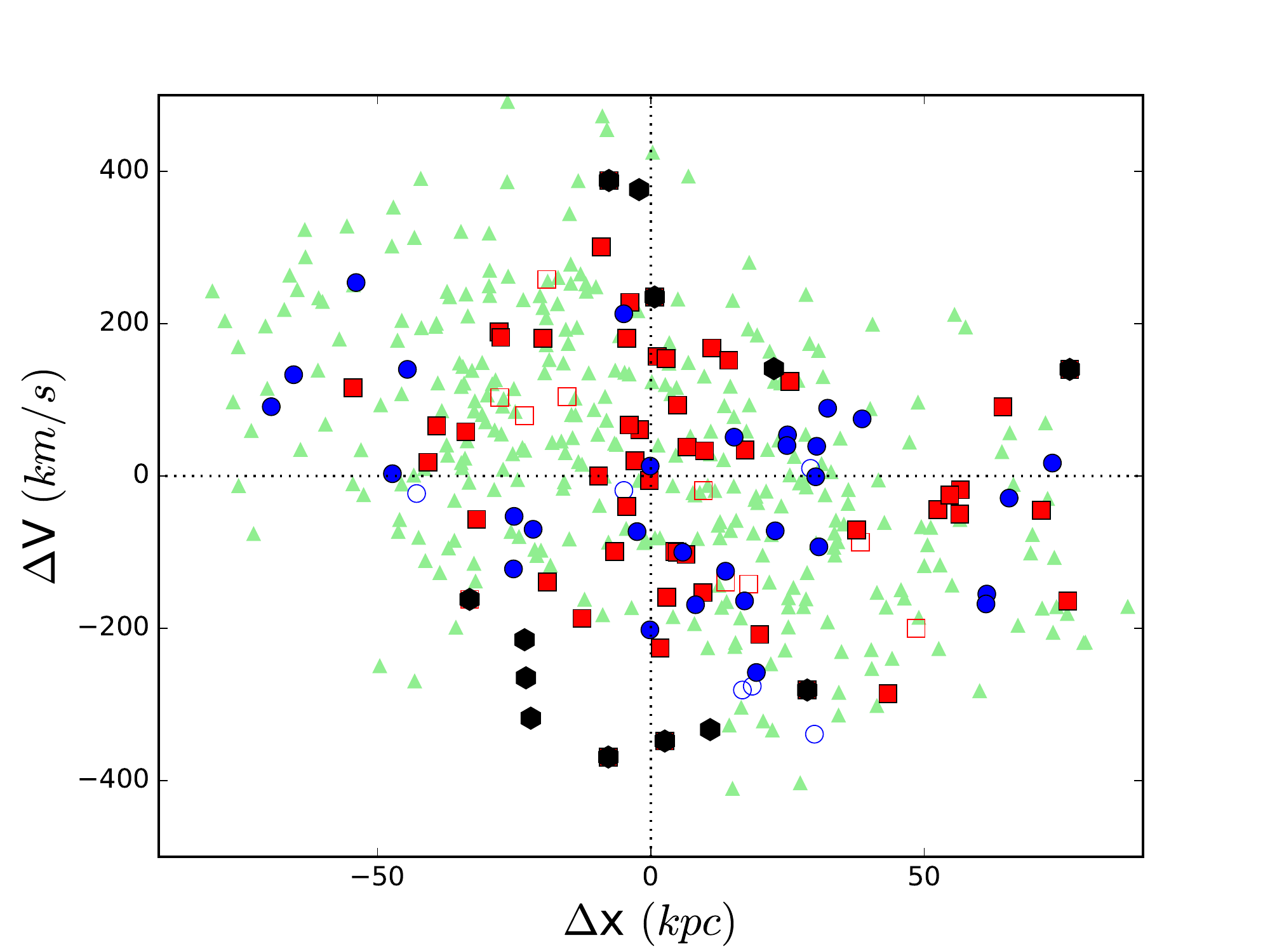}}}\\%
    \subfloat[NGC\,3115]{{\includegraphics[width=0.45\textwidth]{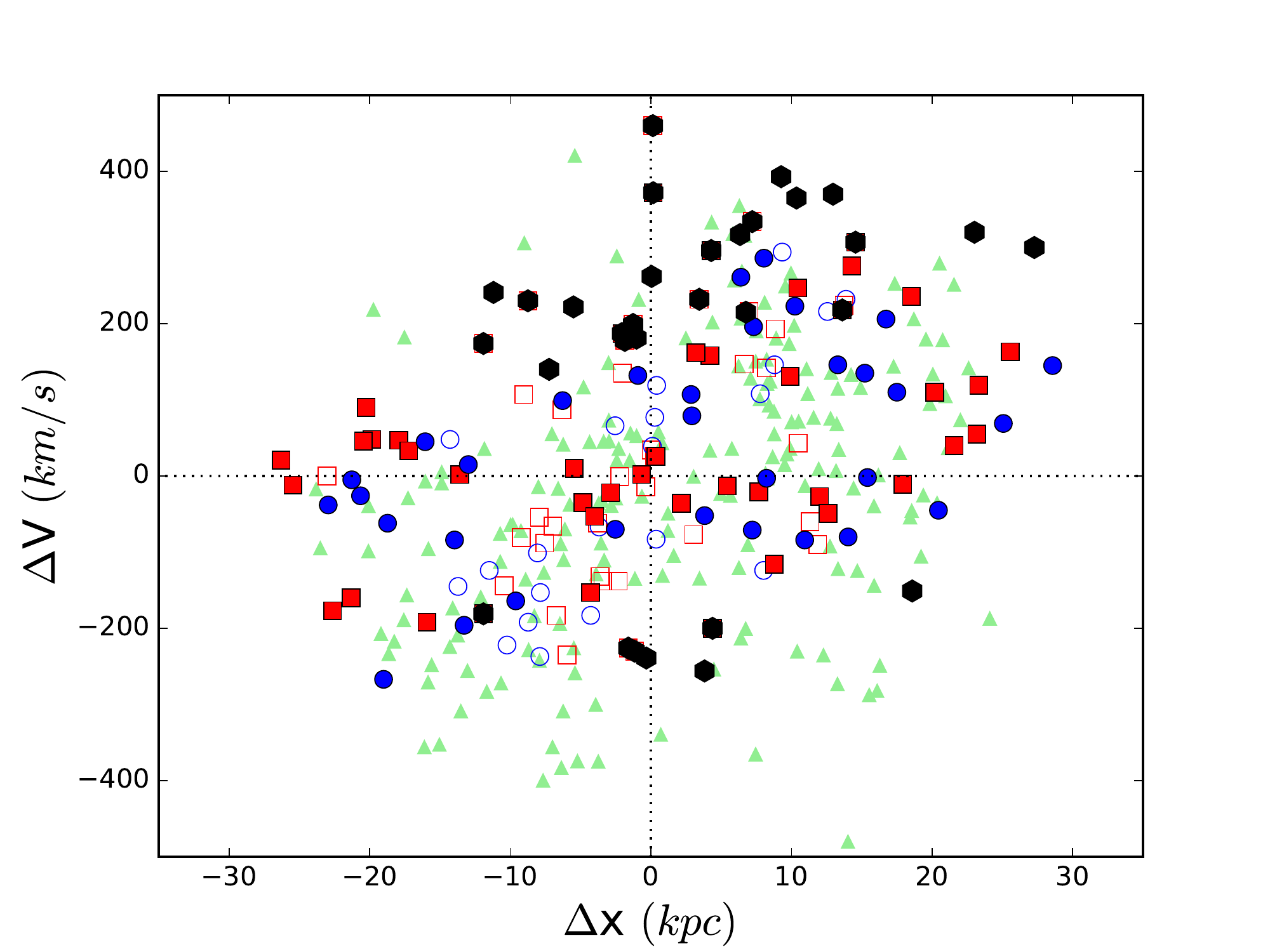}}}\\%
 	\subfloat[NGC\,7457]{{\includegraphics[width=0.45\textwidth]{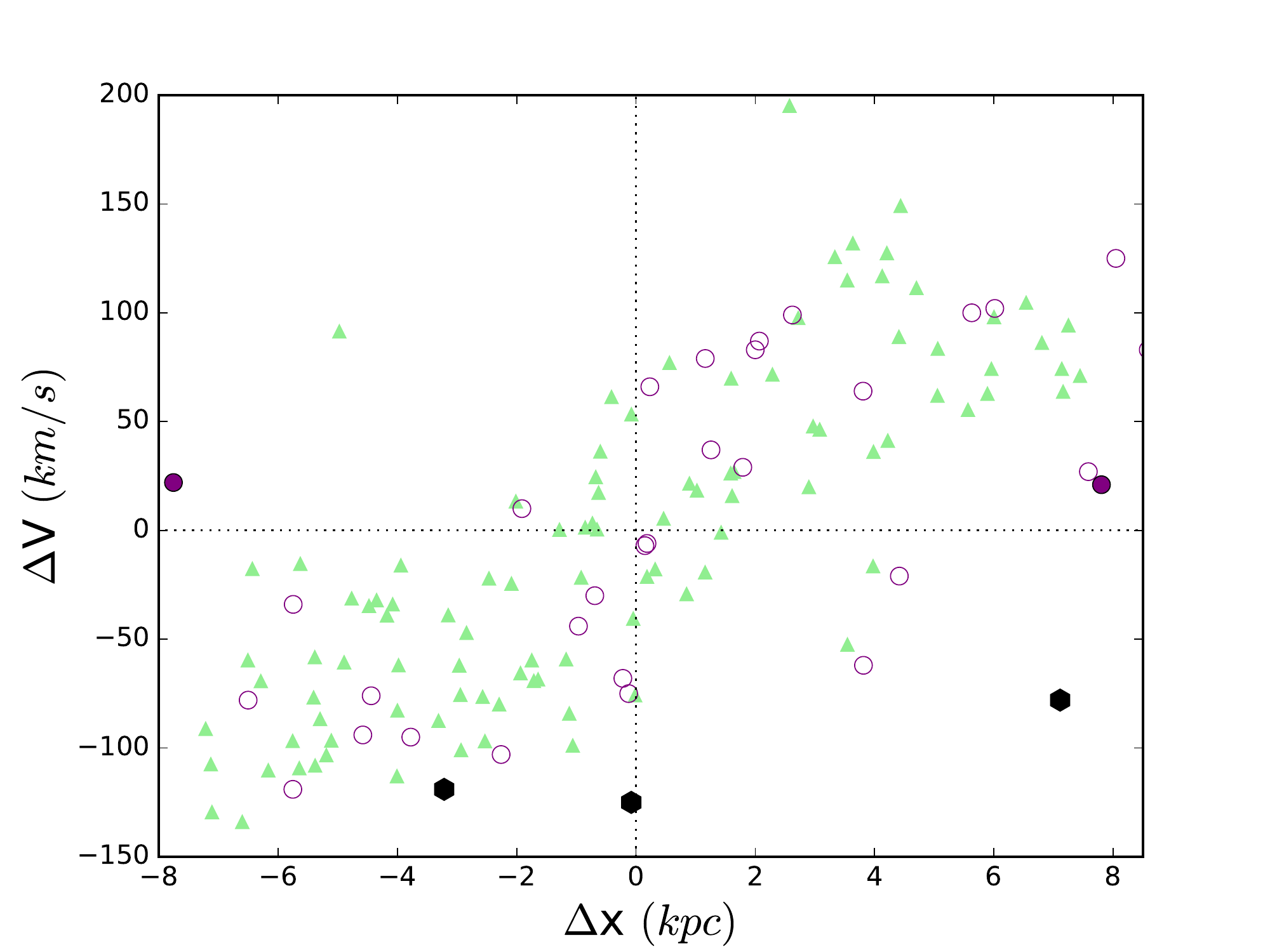}}}%
    \caption{Velocity versus distance along the major axis for GCs and PNe. Markers are the same as in Fig. \ref{fig:phasevel} for all panels. Notice the grouping of rejected GCs in NGC\,3115 around $\delta$ V $\approx$ 300 km/s.}%
    \label{fig:phasecol}%
\end{figure}

It is interesting to combine 1D phase-space diagrams of the GCs with the information obtained from their recovered probabilities of being associated with the disc or the spheroid. In Fig. \ref{fig:phasevel} we use the quantity $\Delta V = V_{los} - V_{sys}$, where $V_{sys}$ is the systemic velocity of a given galaxy, as listed in Table \ref{table:galaxies}, versus galactocentric radius and in Fig. \ref{fig:phasecol} $\Delta x$, the projected distance along the galaxy's major axis. We overplot the PNe for comparison. 

In the 1D phase-space diagrams, GCs that are bound to the system create a 'bell shaped' pattern (see \citet{rocha12}). Spheroid-like objects would lie approximately along the line where $\Delta V =0$, since they have no net rotation, just a natural scatter due to their hot kinematics. On the other hand, disc objects, due to their larger rotational velocities, would show up symmetrically around the galaxy's systemic velocity. Moreover, the broader the distribution around the systemic velocity of the galaxy, the higher the random motions. For instance, objects recently accreted onto the system would generally lie at the outer part of the distribution given that they may not have reached equilibrium yet, with respect to its new host galaxy kinematics (see also \citet{rhee17}).

Fig. \ref{fig:phasevel} shows the 1D phase-space diagrams for NGC\,2768, NGC\,3115 and NGC\,7457. We divide the GCs between disc and spheroid using a value of $L_{Sph}(v_{i}, f_{i})=0.5$. Note that there is a concentration of red GCs towards the centre of NGC\,2768, within 1${R_e}$ (6.37 Kpc) Fig. \ref{fig:phasevel} (a).
However, as seen in Fig. \ref{fig:rcolf} (a), both red and blue sub-populations show high probabilities of belonging to the spheroid. \citet{forbes12} when studying this galaxy kinematics, specifically looking into its red GC population, found that it is related to the kinematics of the galaxy's spheroid. Therefore, it is reasonable to say that, as blue GCs in this galaxy have a high probability of being in the spheroid and are generally not present at the same radii as the red GCs, they are likely located predominantly in the halo. 

In the 1D phase-space diagram for GCs and PNe of NGC\,3115, Fig. \ref{fig:phasevel} (b), one can see a transition from disc-like GCs to spheroid-like GCs with radius. Rejected objects (black markers) from the likelihood analysis also display an interesting distribution in the diagram. Most of them have similar values of $\Delta V$, around 200 km/s.

These GCs might form a group that could have an \textit{ex-situ} origin.

For NGC\,7457, by analysing Fig. \ref{fig:phasevel} (c), we see a more compact distribution of GCs and PNe in the phase-space diagram if compared to panels (a) and (b), with objects showing values of $\Delta V$ $\leq$ 200 km/s. NGC\,2768 and NGC\,3115 have GCs with values of $\Delta V$ in the range of 600 km/s, therefore, NGC\,7457 shows lower velocity dispersion, as expected given the recovered kinematics (see Fig.\ref{fig:like}). GCs and PNe have a high probability of belonging to the disc of the galaxy and share the same loci in this plot. They show a 'U' shaped distribution. This shape might be due to the very low number of GCs and PNe with hot kinematics, consistent with the galaxy being disc dominated \citep{c13b}. This galaxy seems to generally lack a hot kinematic component in its outer regions, beyond around 2 $R_{e}$, quite differently from NGC\,2768 and NGC\,3115 which have, in fact, more prominent bulges and are more massive than NGC\,7457.   

In addition to Fig. \ref{fig:phasevel}, it is interesting to analyse the distribution of the velocities of the tracers versus the projected distance along the major axis of the host galaxies, see Fig. \ref{fig:phasecol}. 
The rejected GCs of NGC\,3115 show interesting properties. In fact, they seem to form a coherent structure, reinforcing what was found in Fig. \ref{fig:phasevel}.

Finally, NGC\,7457 shows a strong disc-like shaped distribution in Fig. \ref{fig:phasecol}, both for PNe and GCs. This was already expected given the rotation profiles in Fig. \ref{fig:like}, where PNe and GCs show similar kinematics.

\section{Discussion}
\label{sec:discussion}

\begin{figure}
	\centering
    \includegraphics[width=9.5cm]{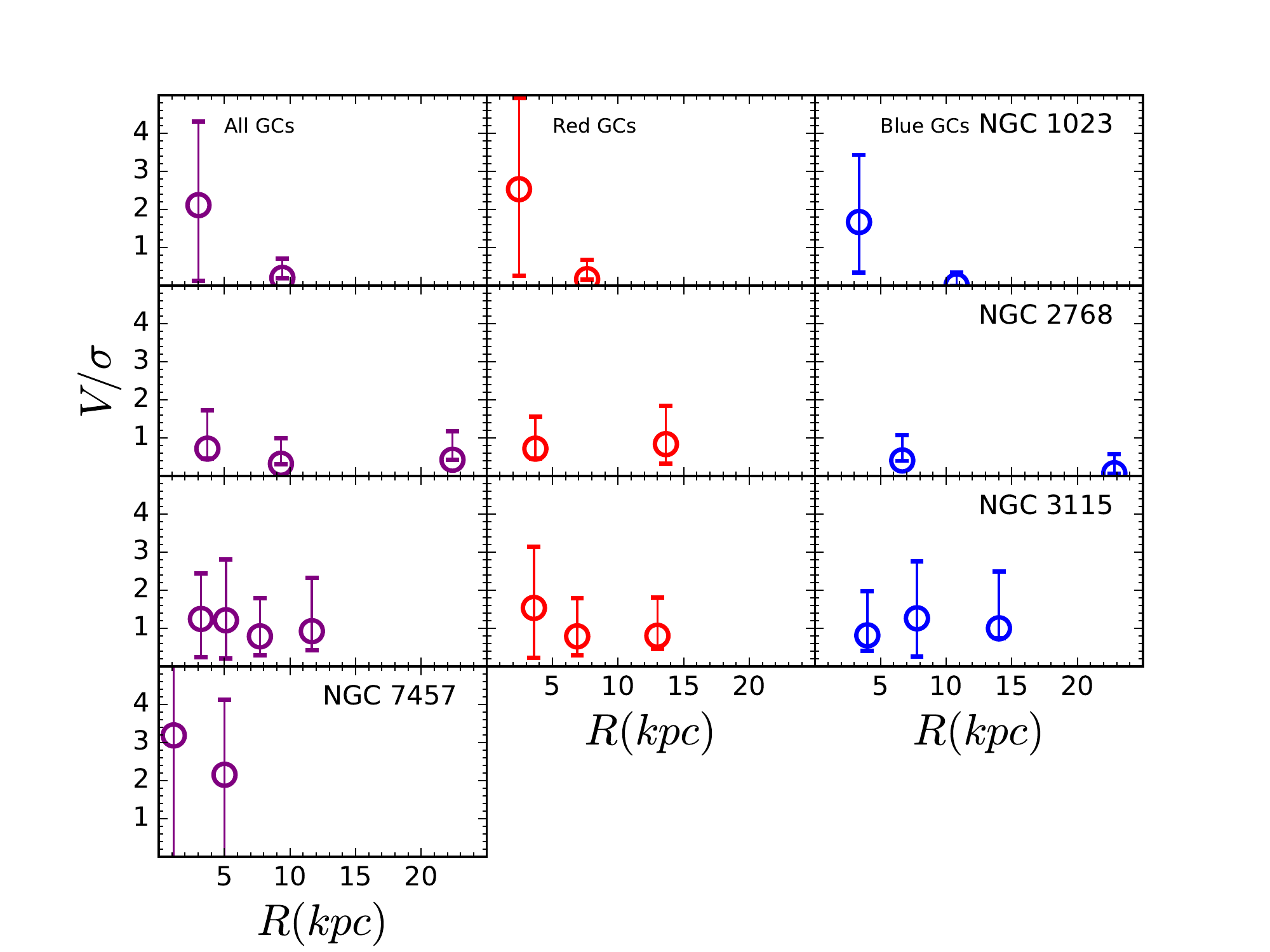}\\%
    \caption{$V/\sigma$ ratio  obtained from GCs, with the addition of NGC\,1023 from \citet{c16}. \textit{Left}, full sample of GCs for each galaxy, \textit{centre}: Red subpopulation of GCs, \textit{right}: blue subpopulation of GCs. NGC\,7457 has no signs of significant bimodality so only one population is shown. The error bars for NGC\,7457 GCs are in the order of $\delta V/\sigma \approx\pm 3.0$, due to the uncertainties in recovering $\sigma$. Note that these values are obtained assuming only one component and not separating disc-like from spheroid-like objects.}%
    \label{fig:vsigma}%
\end{figure}

In this section, we analyse the GC rotation profiles and the $V/\sigma$ ratio of the GCs and we compare our results with simulations. The $V/\sigma$ quantity can be used as an indicator of how much the kinematics of a galaxy's GC system is dominated by rotational velocity, in the case of values higher than 1, or is more influenced by random motions, in the case of a ratio smaller than 1.

\citet{bekki} studied with dissipationless numerical simulations the outcome of various merger scenarios on the GC kinematics of early-type galaxies. They showed that mergers with a proportion of 10:1 are able to produce flattened early-type galaxies, such as lenticulars, and they would impact the kinematics of GCs in such a way that their rotation at large radii would be weaker than at small radii. \citet{bournaurd} showed that minor mergers with a proportion of 4.5:1 would produce a stellar disc with $V/\sigma \approx 1$ for the remnant galaxy and a merger with a proportion of 10:1, such as the ones \citet{bekki} studied, would produce a stellar disc with $V/\sigma \approx 2$. Besides a major merger event \footnote{\citet{bekki} considers mergers with galaxies of masses in the proportion of 4.5:1 as major mergers.}, another likely path on the evolution of galaxies is a sequence of minor mergers. On the other hand, \citet{moody14} showed that multiple minor mergers would not produce fast rotating galaxies \citep{emsellem11}, but instead would be more likely to produce hot, elliptical and slow rotating galaxies. 

In Fig. \ref{fig:vsigma} we show the $V/\sigma$ ratios obtained from the GC sub-populations using the method described in Sec. \ref{sec:methods} and the results from Fig. \ref{fig:like}. One can see that the values of $V/\sigma$ for NGC\,3115 are close to or smaller than 1 for both GC sub-populations and for all radii, which would support the scenario in which this galaxy is a remnant of a merger event where the masses involved follow at least a 4.5:1 proportion. This merger would not have been strong enough to rip the disc structure of the galaxy apart, but would be significant enough to accelerate the gas removal or consumption within the galaxy.

The signs of remnant spiral star forming structures hypothesised by \citep{norris06, guerou} support this formation scenario, where the progenitor of NGC\,3115 was a spiral galaxy that later merged with a smaller companion.
In \citet{c16}, it was argued that for NGC\,1023 the ratio between the rotation velocity and the velocity dispersion in the azimuthal direction is not compatible with a faded spiral galaxy. It was, instead, suggested that NGC\,1023 formed at redshift $\simeq 2$, through the merger of clumps \citep{inoue13}. If one compares Fig.\,6 of \citet{c16} with Fig.\ref{fig:like} of this work it is notable the similarity between the red GCs of NGC\,3115 and NGC\,1023. Hence clumpy disc formation can also be a scenario to explain the kinematics of the red GC population and the higher fraction of GCs with disc-like kinematics in NGC\,3115.
Nevertheless, NGC\,3115 has a higher B/T ratio when compared to NGC\,1023 \citep{c16}. Thus, this galaxy is more consistent with having gone through a series of minor mergers (see also \citet{arnold}).

The results for NGC\,2768 point to a very different scenario compared to the other three galaxies.

\cite{forbes12} argued that the radial distribution of the $V/\sigma$ ratio of red GCs in this galaxy is similar to the one of spheroid PNe and stars, as found in spiral galaxies. Furthermore, we find that for the GCs the $V/\sigma$ ratio is $\leq1$ at all radii, independently of the sub-population considered. According to \citet{bekki} such values of $V/\sigma$ are the result of a 4.5:1 merger, which is consistent with the galaxy having a more prominent spheroidal structure and a high amount of GCs  associated with its spheroid.

Although we are not able to set a single hypothesis for the origin and evolution of NGC\,2768, it is clear from our results that its origin differs from that of NGC\,3115 and NGC\,1023.

NGC\,7457 is a peculiar galaxy, since, unlike the others in the sample, $\sim70\%$ of the GCs are located in the disc. This galaxy has the fewest number of GCs overall (around 210, see \citet{hargis11}). Moreover, we only have spectroscopy for 40 of those, and therefore a slightly less reliable likelihood fit. Nevertheless, the results from the rotation profiles based on GCs are in good agreement with the results for PNe published in \citet*{c13b}, who analysed a sample of 113 such objects. In addition, \citet{hargis11} studied the GC system of NGC\,7457 and concluded that although the spatial distribution of the system is very elliptical, an inclined disc GC population could explain the observations. In the same work, it was suggested that the most likely formation scenario for this galaxy is through a merger event involving galaxies with unequal masses. NGC\,7457 is a field galaxy, so gas stripping mechanisms, which are usually related to dense environments, are very unlikely to be major contributors here. Even so, it could be the result of starvation, where the isolated galaxy, after consuming all its gas reservoir, slowly stops creating young stars \citep{bekki02}. Presently, in the literature, scenarios for early-type galaxy formation involving a two-phase evolution are gaining popularity \citep{oser10}. In these scenarios, the bimodality of GCs in colour, metallicity and age would be a more natural outcome. \citet{hargis11} and \citet{pota13} were not able to detect bimodality in colour for the GCs in this galaxy (but see \citep{peacock17}).

The $V/\sigma$ ratio for this galaxy is more than 2, which would mean, following \citet{bournaurd}, that at most it has undergone a merger event with a proportion of around 10:1. Moreover, the simulations of \citet{bekki} indicate that the GC systems of galaxies that suffered minor mergers, as the one that would form flattened disc galaxies, such as NGC\,7457, should retain a more spherical structure with little rotation at larger radii, which does not seem to be the case here. 
Another piece of information that can be added to this puzzle comes from \citet{alabi17}. In this work, the dark matter fraction at large radii for NGC\,7457 was found to be around 0.9, within 5 $R_{e}$. This is greater than the dark matter fraction values found for NGC\,2768, NGC\,3115 and NGC\,1023 which are all around 0.6. With this information, \citet{alabi17} calculated an assembly epoch for the halo of NGC\,7457 at $z\approx4.4\pm1.1$, or 12.3 Gyr ago.
However, the mean luminosity-weighted ages for the stellar content of the central regions of this galaxy and of some of its GCs are around 3-7 Gyr \citep{sil02,cho08,mcdermid15}. Therefore, the assembly of the halo of the galaxy would have taken place long before some of the GCs and stars at the centre of the galaxy were created. This formation scenario is compatible with a secular evolution scenario, with no mergers at least after $z\simeq4$. However, clumpy disc formation can also explain the fact that NGC\,7457 has a disc dominated GC population \citep{inoue13}.

\begin{figure}
	\centering
    \subfloat[]{{\includegraphics[width=8cm]{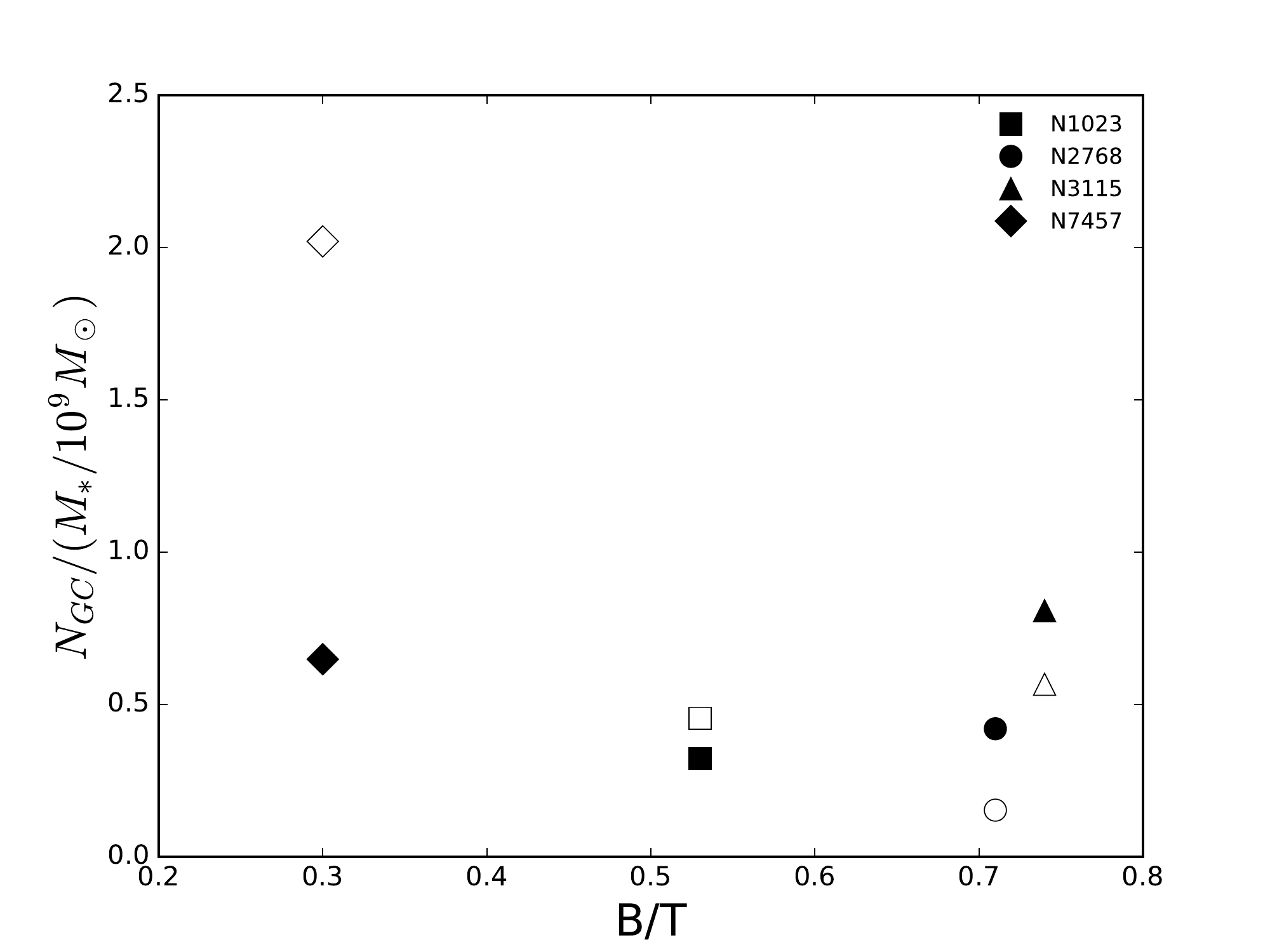}}}\\%
    \subfloat[]{{\includegraphics[width=8cm]{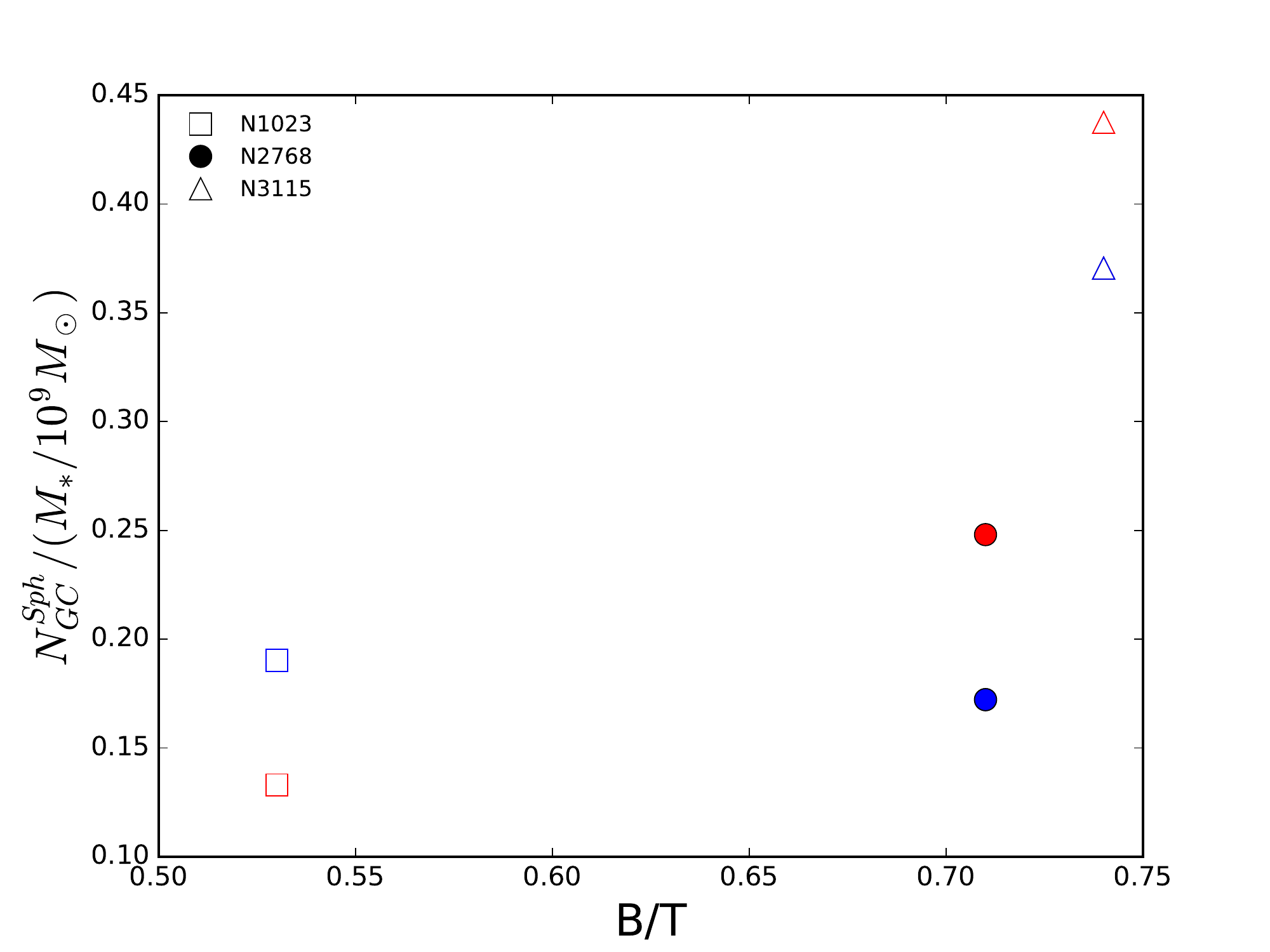}}}%
    \caption{\textit{Upper panel:} Comparison between the number of GCs in the spheroid (filled markers) and the disc (open markers) of our sample galaxies, divided by total stellar mass ($M_{\text{*}}$) and the bulge-to-total ratio of the galaxy light profile, as obtained in \citet*{c13b}.\textit{Lower panel:}Same as the upper panel, but only accounting for the spheroidal GC population of all galaxies, with the exception of NGC\,7457, and separated in red and blue colour subpopulations. }%
    \label{fig:miscplots}%
\end{figure}

In Fig. \ref{fig:miscplots} (a), we compare the number of GCs associated with the spheroid or disc components of its host galaxy, normalised by total stellar mass, with the bulge-to-total (B/T) ratio of the host galaxy light profile from \citet{c13b}. We do not find that the number of GCs increases with the host galaxy B/T ratio. Nevertheless, if 

we consider only the number of GCs that belong to the spheroid (in panel b\footnote{NGC\,7457 is left out due to is negligible amount of spheroidal GCs, when compared to the other three galaxies.}), we can see a trend with B/T. This suggests that the number of spheroidal GCs in a given galaxy can be estimated by its photometric properties, such as the fraction of bulge light, but disc GCs, on the other hand, do not follow a similar relation. This discrepancy with disc GCs then impacts the total number of GCs of a given galaxy when compared to its B/T ratio and should not be ignored.
On a final note, Fig. \ref{fig:miscplots} (a) also showcases the unusual proportion of GCs in NGC\,7457 in proportion to its mass when compared to the other galaxies.

\section{Summary and Conclusions}
\label{sec:conclusion}

In this work we recovered the kinematics of the GC systems of three lenticular galaxies in low density environments: NGC\,2768, NGC\,3115 and NGC\,7457. We employed the method presented in \citet{c16} for NGC\,1023 to obtain the probability of every GC to belong to the disc or the spheroid of the galaxy, using PNe kinematics and K-band photometry. The results point to a different formation scenario for each galaxy. NGC\,2768 hosts a very prominent spheroidal GC population. NGC\,3115 has the same number of GCs associated with the disc and the spheroid components. NGC\,7457, interestingly, at least for the sample used in this work, has the majority of its GCs compatible with disc-like kinematics. 

Following the simulations of \citet{bekki} and \citet{bournaurd}, and the fact that all galaxies in this sample are isolated or in small groups, we investigated the possibilities of merger origins for the galaxies studied. Although such hypothesis remains reasonable, the kinematics of red GCs associated with the disc component are compatible with clumpy disc formation. In the case of NGC\,7457, our results seem to point towards a secular evolution from a regular spiral galaxy, or clumpy disc formation. Also of interest, the GC colour sub-populations of NGC\,2768  display distinct kinematic behaviours, a feature not present in NGC\,3115, NGC\,7457 or even NGC\,1023 from \citet{c16}. For all GC systems there is no clear correlation between the component of the galaxy they are likely to belong to and their colour. We find also a population of GCs in NGC\,3115 which are not likely to belong to any of the modelled components, within a confidence interval of 2.3$\sigma$. Those GCs could be related to a recent accretion event or to components not explicitly included in our model, such as the halo. This is to be further investigated.

In summary, this work shows that the structure and kinematics of lenticular galaxies in low density environments is very diverse and more complex than expected by most formation scenarios proposed in the literature.

\section*{Acknowledgements}
We thank Basilio Santiago, Hor\'acio Dottori, Caroline Foster, Jos\'e Eduardo Costa, Vincenzo Pota and Lo\"ic Le Tiran for interesting discussions and contributions. We also thank the anonymous referee for valuable comments. EZ, ACS and CMdO acknowledges funding from CNPq. ACS acknowledges funding from CNPq-403580/2016-1 and 310845/2015-7, PqG-FAPERGS-17/2551-0001 and L\textquotesingle Or\'eal UNESCO ABC Para Mulheres na Ci\^encia. CMdO acknowledges funding from FAPESP (process number 2009/54202-8). DF thanks the ARC for financial support via DP160101608. AJR was supported by National Science Foundation grant AST-1616710 and as a Research Corporation for Science Advancement Cottrell Scholar. 

% The Acknowledgements section is not numbered. Here you can thank helpful
% colleagues, acknowledge funding agencies, telescopes and facilities used etc.

%%%%%%%%%%%%%%%%%%%%%%%%%%%%%%%%%%%%%%%%%%%%%%%%%%



\bibliographystyle{mnras}
\bibliography{references.bib} % if your bibtex file is called example.bib


%%%%%%%%%%%%%%%%%%%%%%%%%%%%%%%%%%%%%%%%%%%%%%%%%%

%%%%%%%%%%%%%%%%% APPENDICES %%%%%%%%%%%%%%%%%%%%%

%%%%%%%%%%%%%%%%%%%%%%%%%%%%%%%%%%%%%%%%%%%%%%%%%%

% Don't change these lines
\bsp	% typesetting comment
\label{lastpage}
\end{document}